\pdfoutput=1
\documentclass[11pt,a4paper]{article}

\usepackage[utf8]{inputenc}
\usepackage[T1]{fontenc}
\usepackage{amsmath,amssymb,amsthm,mathtools}
\usepackage{physics}
\usepackage{bm}
\usepackage[dvipsnames,svgnames,x11names]{xcolor}
\usepackage{geometry}
\usepackage{hyperref}
\hypersetup{colorlinks=true,linkcolor=blue!60!black,
            citecolor=green!50!black,urlcolor=blue!70!black}
\usepackage{enumitem}
\usepackage{booktabs}
\usepackage{graphicx}
\graphicspath{{./}}
\usepackage[section]{placeins}
\usepackage{etoolbox}
\usepackage{cleveref}
\usepackage{tikz}
\usetikzlibrary{decorations.pathmorphing,arrows.meta,positioning,
                calc,shapes.geometric}

\tikzset{
  xprop/.style={
    decorate,
    decoration={snake, amplitude=1.2mm, segment length=4mm,
                post length=1mm, pre length=1mm},
    thick
  },
  xprop dressed/.style={
    double, double distance=2pt,
    decorate,
    decoration={snake, amplitude=1.2mm, segment length=4mm,
                post length=1mm, pre length=1mm},
    very thick, blue!60!black
  },
  yprop/.style={thick, dashed},
  vtx/.style={circle, fill=black, inner sep=2pt},
  vtx3/.style={regular polygon, regular polygon sides=3, fill=red!70!black,
               inner sep=1.5pt},
  vtx4/.style={regular polygon, regular polygon sides=4, fill=green!50!black,
               inner sep=2pt},
}

\newcommand{\EE}{\mathbb{E}}
\newcommand{\PP}{\mathbb{P}}

\newcommand{\Cov}{\operatorname{Cov}}
\newcommand{\ii}{\mathrm{i}}
\newcommand{\calS}{\mathcal{S}}

\newcommand{\Veff}{V_{\mathrm{eff}}}
\newcommand{\thetaeff}{\theta_{\mathrm{eff}}}
\newcommand{\supp}[1]{Technical Supplement, #1}
\newcommand{\timingnote}{Timings are indicative only: a single-thread Python prototype in
double precision (NumPy, with Numba-compiled kernels for the conditional-mode closures),
quoted for orders of magnitude rather than production performance.}

\title{\textbf{Dyson--Schwinger Effective-Action Methods\\[4pt]
for Rough Volatility}\\[8pt]
\large A Correlation--Response Architecture for Calibration, Exotics and Risk}
\author{Frédéric Pauquay\thanks{Correspondence: \texttt{fpauquay@DRWHoldings.com}.}\\[2pt]
\normalsize Global Quantitative Modelling \& Analytics, DRW}
\date{}

\begin{document}
\maketitle

\begin{abstract}
We develop a non-perturbative framework for stochastic-volatility option pricing
built on the two-particle-irreducible (2PI) effective action and the
Dyson--Schwinger gap equations of quantum field theory. In log-price,
log-volatility or Lamperti coordinates, which remove as much state dependence from
the diffusion coefficients as possible, the joint law of the state variables is
approximated by a self-consistent Gaussian. Its mean and its effective diffusion
matrix (the Gaussian average of the instantaneous covariance, i.e.\ the noise
self-energy of the 2PI action) follow from the 2PI stationarity conditions, while
its effective drift Jacobian is the Gaussian average of the drift gradient
(statistical linearisation). At this order the equal-time system coincides with
Gaussian assumed-density filtering; the 2PI view supplies what comes next. For the
exponential vertices of lognormal volatility the two-particle-irreducible skeletons
are available in closed form: two vertices exchanging $n\ge3$ lines sum to the melon
kernel $e^{G}-1-G-\tfrac12G^2$, which contains the sunset, the basketball and every
higher melon, and a correlated leverage term is brought to the same local form by
It\^o's formula. The resulting self-energy is non-local in time even for a Markovian
model. In exp-OU the evaluated melon level cuts the Hartree error by $20$--$130\times$,
to $0.01$--$0.11$\,bp at $\rho\in\{0,\pm0.4,-0.7\}$; on the rough-Bergomi Volterra field
at $\rho=0$ it reaches $0.1$--$1.2$\,bp in mild and moderate regimes. Whether the
dressed inverse propagator is local in time separates Markovian models, where the Hartree gap equation collapses to a few
ODEs, from rough (Volterra) models, where the full two-time propagator is retained.
At nonzero correlation the field is integrated deterministically by conditioning on its
two leading Gaussian modes: given the modes the field is still Gaussian, the moments of
the composite pair (integrated variance, leverage integral) are exact finite sums, and a
moment-matched law for that pair closes the problem at each mode node. For the CEV
models the same modes feed the exact absorbed-CEV propagator, the correlated
displacement, which acts as a drift along the variance clock, enters through a Dyson
equation around that propagator, and the equation is resummed by composing the
propagator across jumps at Chebyshev--Lobatto clock nodes. Against Monte Carlo
references these conditional-mode closures agree with a $2^{24}$-path rough-Bergomi
reference approximately at its resolution ($0.01$--$0.1$\,bp, down to $H=0.07$, $\rho=-0.9$), price
SABR to a mean of $0.97$\,bp over a $72$-cell grid with maturities to ten years
($1.06$\,bp on held-out cells, against $75$\,bp for Hagan's expansion), and rough SABR
to $0.92$\,bp over a $72$-cell grid, with exp-OU at $0.1$--$0.2$\,bp and rough Heston as an
exactly-transformable control. The same correlation--response architecture then
supplies exotics and risk: conditional on the volatility field, forward-start
smiles and continuously-monitored barriers reduce to field-only quadratures---in
rough Bergomi directly on the native Volterra field, with no Markovian lift or
spot-path simulation---and the causal response block, which for a free field is the
Volterra kernel itself, yields the latent-factor impulse-vega curve in one adjoint
contraction. Complete derivations, extended benchmark grids and fuller
treatments of the secondary applications (stochastic-rate FX local volatility,
quadratic Gaussian Volterra variance, and the arbitrage-free FX triangle) are
provided in an accompanying technical supplement available as an ancillary file on
the arXiv abstract page.
\end{abstract}

\tableofcontents
\bigskip

\section{Introduction}
\label{sec:intro}

Stochastic-volatility pricing is, at its core, the problem of computing an
expectation of a nonlinear functional of a Gaussian field. In log-price and
log-volatility coordinates the driving noise is Gaussian; the smile is generated
entirely by the nonlinearities that couple the fields---the exponential map from
log-variance to variance, the CEV power $F^\beta$, the square-root of a variance
process---and by the memory of the volatility kernel. The computational target is
therefore the law (in practice the first few cumulants) of a Gaussian field seen
through a nonlinear map, and the design choice that determines everything else is
how that map is handled: expanded in powers of the fluctuation, as the asymptotic
methods do, or kept intact.

The models that resist closed-form treatment are exactly the non-affine ones.
Affine models are the easy case: Heston~\cite{Heston1993} and, remarkably, rough
Heston~\cite{ElEuchRosenbaum2019} linearise the problem, so their characteristic
function follows from a (fractional) Riccati equation. That structure is fragile.
The moment the variance becomes a nonlinear function of a latent Gaussian factor
($v=e^X$ in exp-OU and rough Bergomi, or the CEV map $F^\beta$ in SABR and rough
SABR), the Riccati equation is lost, and with it the closed form. These are the
models we target, and they share one difficulty: the nonlinearity has no small
parameter, so a low-order Taylor approximation is accurate only while the
fluctuation is small and fails in precisely the long-dated, high-vol-of-vol regime
that matters most. The workhorse tools sit at two extremes. The asymptotic
expansions---Hagan's SABR formula~\cite{Hagan2002}, Paulot's second-order implied
volatility~\cite{Paulot2015}, the heat-kernel expansions of
Lorig--Pagliarani--Pascucci~\cite{LPP2017}---are fast and closed-form but
perturbative, losing accuracy once $\nu^2 T\gtrsim 1$. Monte Carlo and Markovian
lifting~\cite{AbiJaberElEuch2019} are accurate and general but slow, and lifting
adds a kernel-truncation error that is hard to control for very rough kernels.
What is missing is a method in between: non-perturbative, so that it stays
accurate well beyond the short-time and small-vol-of-vol regime, yet deterministic
(or, where the field average is done by randomised quasi-Monte Carlo, of small and
measured scramble variance) and fast enough for calibration.

We supply such a method by importing the Dyson--Schwinger, or 2PI,
effective-action formalism of Cornwall, Jackiw and
Tomboulis~\cite{CJT1974,Berges2004,Roberts1994}. The 2PI action is a functional of
the mean field and the propagator (the covariance); its stationary point yields a
self-consistent \emph{gap equation} for the propagator. Truncating the action at a
given loop order does not truncate the series in the coupling---it resums an
infinite subclass of diagrams into each retained skeleton, and this is what makes
the approximation non-perturbative. The simplest truncation is the Hartree
(one-loop) approximation: a Gaussian ansatz for the fluctuations, but with the
interaction evaluated through the \emph{exact} Gaussian moment-generating function
$\EE[e^{a\phi}]=e^{a\bar\phi+a^2\gamma/2}$ rather than a Taylor cut of it. That
single identity resums the one-loop tadpole and the entire daisy tower; the
non-local skeletons (sunset, basketball) are distinct graphs, and for exponential
vertices they too sum in closed form (\Cref{sec:diagrams}). Propagated forward in time
as an equal-time Gaussian, in the manner of assumed-density filtering, the Hartree
level gives the scheme we call the Self-Consistent Gaussian (SCG).

\paragraph{Related work.}
A field-theoretic view of derivative pricing is not itself new: path-integral
formulations date to Baaquie's quantum-finance programme~\cite{Baaquie2004}, and
nonlinear-PDE and McKean--Vlasov techniques are surveyed by Guyon and
Henry-Labord\`ere~\cite{GuyonHL2012}. The closest methodological antecedent, however,
lies in nonequilibrium physics rather than finance: the 2PI effective action for
\emph{classical} stochastic dynamics, built on the
Martin--Siggia--Rose--Janssen--De\,Dominicis response-field construction and closed by
exactly this Hartree self-consistency in the equal-time covariance, is standard
there~\cite{MartinSiggiaRose1973,Janssen1976,DeDominicisPeliti1978,Kamenev2011}.
The single closest antecedent is Bode~\cite{Bode2022}, who combines the 2PI action with
the MSR construction to derive self-consistent equations for the first two cumulants of
a nonlinear classical SDE, and already names geometric Brownian motion and the Heston
model as prospective targets; the construction below runs parallel to his on the physics
side. We do not claim that machinery as ours. At the Gaussian level the equal-time
equations are those of Gaussian assumed-density filtering and statistical
linearisation~\cite{Sarkka2013}, and of the equal-time limit of~\cite{Bode2022}; the
comparison is made term by term in \supp{S2}. What the 2PI construction adds beyond
those methods is the next order. Bode's cumulant equations and assumed-density filters
close the hierarchy by truncating moments; the 2PI action instead keeps the exponential
vertex intact and organises the corrections as two-particle-irreducible skeletons in the
dressed propagator. For exponential vertices these skeletons have closed forms
(\Cref{sec:diagrams}), so the hierarchy can be evaluated rather than only named, and its
first correction is a non-local self-energy that no equal-time filter contains.
Our contribution on the method side is this transfer to option pricing and rough
volatility, with the vertex evaluated through the exact Gaussian moment-generating
function, the skeleton corrections evaluated in closed form and, for the CEV models, a
Dyson equation around the exact absorbed-CEV propagator resummed across clock jumps; on the application
side it is the observation that the same two-time objects serve calibration, exotics
and risk.

\paragraph{A field-theory-to-finance dictionary.}
For readers less fluent in effective actions, the correspondence used throughout
is compact (\Cref{tab:glossary}): each field-theory object is a familiar financial
or statistical quantity, and no step below needs more than this dictionary.

\begin{table}[h]
\centering
\small
\begin{tabular}{@{}ll@{}}
\toprule
\textbf{Field-theory object} & \textbf{Finance / statistics meaning} \\
\midrule
Effective action $\Gamma[\bar\phi,G]$ & variational free energy; Legendre transform of the log-CF \\
Mean field $\bar\phi$ & forward / drift-adjusted level of the state \\
Propagator $G$, correlation $C(t,s)$ & term structure of state-variable covariances \\
Dressed $G$ vs.\ bare $G_0$ & covariance with / without self-consistent corrections \\
Self-energy $\Sigma$ & covariance correction from the nonlinearity \\
Gap / Dyson equation $G^{-1}=G_0^{-1}-\Sigma$ & self-consistency for the covariance \\
Tadpole / daisy term & Jensen convexity gap, $\EE[e^{X}]$ vs $e^{\EE[X]}$ \\
Melon skeletons (sunset, basketball, \dots) & non-local covariance of two vertex insertions \\
Response propagator $R(t,s)$ & Green's function $=$ sensitivity (a Greek) \\
Martingale normalisation $\varphi(-\ii)=1$ & no-arbitrage constraint on the drift \\
Instanton / freezing & rare large-variance (heavy-tail-dominated) regime \\
\bottomrule
\end{tabular}
\caption{The field-theory-to-finance dictionary used throughout.}
\label{tab:glossary}
\end{table}

The results fall into three groups, and we keep them apart throughout. The first is
the closure proper: in the Markovian non-affine models (exp-OU, the fast SABR stack, the
FX local-volatility spot) the prices come from the Gaussian gap equation, and in exp-OU and on the
rough-Bergomi field at zero correlation from its evaluated skeleton corrections. The
second is conditional-field pricing at nonzero correlation. Conditional on the
volatility field (Romano--Touzi) the problem is exactly Gaussian, or exactly absorbed
CEV up to a correlated drift, and the field average is done either by PCA-ordered
quasi-Monte Carlo or deterministically by the conditional-mode closures: the two leading
Gaussian modes of the free field are integrated by quadrature, the moments of the
composite pair (integrated variance, leverage integral) given the modes are exact, and a
moment-matched law for that pair replaces the remaining field integral. For SABR and
rough SABR the correlated drift is then resummed around the exact absorbed-CEV
propagator by jumps at Chebyshev--Lobatto clock nodes. These closures price rough
Bergomi, SABR and rough SABR to about a basis point or better, deterministically, and
the rough-Bergomi forward smile likewise; the quasi-Monte Carlo field average is kept for
rough barriers. The ingredients of this second group differ in status. Conditioning on
the field is the evaluation device the framework relies on at nonzero correlation, whether
the field average is then done by quadrature over its modes or by quasi-Monte Carlo, and
the composite node law is a moment-matched closure of that conditional problem rather than
a skeleton of the action. The jump closure is derived: it is the Dyson equation for the
correlated drift around the exact absorbed-CEV propagator, its jump sizes are the
first-order Duhamel weights of that equation, so the first-order term is reproduced, and
composing the exact propagator across the jumps carries the higher orders. The third group is structural: the marginal
characteristic function, the exotic conditional law and the risk response are three
contractions of one pair of two-time objects---the correlation and the causal
response. On the implementation side the work is a conditional-Gaussian engine that integrates out as much
randomness as each problem allows and reuses the identical covariance and response
tensors across calibration, path-dependent pricing and sensitivities; in the rough
branch it runs directly on the native two-time covariance of the Volterra field,
with no Markovian lift and no spot-path simulation, so forward-start smiles and
continuously-monitored barriers become field-only quadratures amortised across
strikes. What the framework produces is therefore specific: the Gaussian gap equations,
the skeleton hierarchy evaluated in closed form, and the Dyson resummation around exact
reference propagators, each with a truncation that can be raised. Other pieces are
established technology that it uses as evaluation devices and does not derive
(conditional Monte Carlo for rough Bergomi~\cite{McCrickerdPakkanen2018}, exact CEV
transition densities, martingale re-centring, pathwise/adjoint sensitivities), and we do
not claim them. The practical counterpart is that one implementation spans the family.
Complete derivations, extended benchmarks and the fuller technical treatment of the
secondary applications are deferred throughout to the accompanying technical
supplement; the main text keeps everything needed to evaluate and cite the method,
including the secondary applications themselves in condensed form. The whole
architecture---one field, three propagator blocks, two structural
regimes---is summarised in \Cref{fig:architecture}.

\begin{figure}[t]
\centering
\begin{tikzpicture}[
  font=\small, >=Stealth, node distance=6mm,
  box/.style={draw, rounded corners, align=center, inner sep=4pt,
              minimum height=8mm, minimum width=20mm},
  prop/.style={draw, rounded corners, align=center, inner sep=4pt,
               fill=blue!6, minimum height=8mm, minimum width=24mm},
  deliv/.style={draw, rounded corners, align=center, inner sep=4pt,
              fill=green!6, minimum height=8mm, minimum width=26mm}
]
  \node[box] (model) {model SDE};
  \node[box, right=9mm of model] (field) {self-consistent\\Gaussian field};
  \node[prop, right=16mm of field] (poff) {off-diagonal\\$C(t,s)$};
  \node[prop, above=5mm of poff] (peq) {equal-time\\$C(t,t)$};
  \node[prop, below=5mm of poff] (pr) {causal response\\$R(t,s)$};
  \node[deliv, right=13mm of peq] (dv) {vanilla smiles,\\calibration};
  \node[deliv, right=13mm of poff] (df) {forward smiles,\\barriers};
  \node[deliv, right=13mm of pr] (dg) {latent-factor\\Greeks};
  \draw[->] (model) -- (field);
  \draw[->] (field) -- (peq);
  \draw[->] (field) -- (poff);
  \draw[->] (field) -- (pr);
  \draw[->] (peq) -- (dv);
  \draw[->] (poff) -- (df);
  \draw[->] (pr) -- (dg);
  \node[align=center, font=\footnotesize\itshape, anchor=north]
    at ([yshift=-7mm]pr.south -| field)
    {Markovian: equal-time ODEs \quad$\vert$\quad Rough: full two-time covariance};
\end{tikzpicture}
\caption{The correlation--response architecture. A model SDE is mapped to a
self-consistent Gaussian volatility field; the effective action then supplies three
two-time blocks at no extra calibration cost---the equal-time correlation $C(t,t)$
(vanilla smiles and calibration, Part~I), the off-diagonal correlation $C(t,s)$
(forward smiles and barriers, Part~II) and the causal response $R(t,s)$ (latent-factor
Greeks, Part~III). The single structural fork---whether the dressed inverse propagator
is local in time---separates the Markovian regime (the gap equation collapses to
equal-time ODEs) from the rough regime (the full two-time covariance is retained).}
\label{fig:architecture}
\end{figure}

\section{The Dyson--Schwinger Framework}
\label{sec:framework}

\subsection{State variables, path measure and the effective action}
\label{sec:framework-action}

Pricing a European claim reduces to the characteristic function
$\varphi(u)=\EE[e^{\ii u X_T}]$ of the log-return $X_T=\log(S_T/S_0)$, from which
every strike follows by Fourier inversion; the task is to compute the low
cumulants of the state variables without linearising the dynamics that generate
the smile. Collect the state in $\bm\phi(t)=(\phi_1,\dots,\phi_n)$, with
$\dd\phi_a=\mu_a(\bm\phi)\,\dd t+\sum_c\sigma_{ac}(\bm\phi)\,\dd W_c$, and choose
coordinates (log-price, log-volatility, or the Lamperti variable) that remove as
much state dependence from the diffusion as possible, so that the nonlinearity sits
mainly in the drift. Where the diffusion is constant the path law is of Boltzmann
form, $\propto e^{-\calS[\bm\phi]}$, with the quadratic Onsager--Machlup action
\begin{equation}\label{eq:action}
  \calS[\bm\phi]=\tfrac12\int_0^T(\dot{\bm\phi}-\bm\mu)^{\!\top}
  \mathbf\Sigma_0^{-1}(\dot{\bm\phi}-\bm\mu)\,\dd t,
  \qquad \mathbf\Sigma_0=\sigma\sigma^{\!\top},
\end{equation}
in the pre-point (Itô) discretisation. A residual state dependence---the
volatility factor multiplying the Lamperti noise in SABR, or
$\sigma_{\mathrm{loc}}(t,S)$ in local volatility---is carried by the field-dependent
noise vertex of the response-field form of \Cref{sec:framework-response}, and
$\mathbf\Sigma_0$ in \eqref{eq:action} is then $\sigma\sigma^{\!\top}$ evaluated at the
mean $\bar{\bm\phi}$ (\supp{S1}).
Splitting $\calS=\calS_0+\calS_{\text{int}}$
into the linearised Gaussian dynamics and the nonlinearity is the only structural
input the method requires. Tilting the measure by a source and forming
$W[\mathbf J]=\log\EE[\exp\int\mathbf J\cdot\bm\phi]$, the derivatives of $W$ are
the cumulants: the mean $\bar\phi_a=\delta W/\delta J^a$ and the two-time
covariance $G^{ab}(t,s)=\delta^2 W/\delta J^a_t\delta J^b_s$. The characteristic
function is the generating functional at a terminal imaginary source on the
log-price, $J_X(t)=\ii u\,\delta(t-T)$. The mean and covariance alone do not fix the
smile; within the Gaussian closure they are the inputs from which the pricing
transform is built, by solving the closure under that frequency-dependent source or
by averaging conditionally Gaussian laws over the volatility field.

The route that survives strong nonlinearity is not to expand $W$ in powers of
$\calS_{\text{int}}$ but to Legendre-transform it in two sources at once---one
conjugate to the mean, one bilocal source conjugate to the covariance---giving a
functional of the mean and covariance directly,
\begin{equation}\label{eq:Gamma2PI}
  \Gamma[\bar{\bm\phi},\mathbf G]=\calS[\bar{\bm\phi}]
  +\tfrac12\operatorname{Tr}\ln\mathbf G^{-1}
  +\tfrac12\operatorname{Tr}(\mathbf G_0^{-1}\mathbf G-\mathbf 1)
  +\Phi[\bar{\bm\phi},\mathbf G],
\end{equation}
with $\mathbf G_0^{-1}=\delta^2\calS_0/\delta\bm\phi^2$ and $\Phi$ carrying
$\calS_{\text{int}}$. This representation is exact, and its content is that the true
mean and covariance are its stationary point, $\delta\Gamma/\delta\bar{\bm\phi}=0$
and $\delta\Gamma/\delta\mathbf G=0$. The second condition is the Dyson equation
\begin{equation}\label{eq:dyson}
  \mathbf G^{-1}(t,s)=\mathbf G_0^{-1}(t,s)-\bm\Sigma[\mathbf G](t,s),
  \qquad \Sigma^{ab}(t,s)=-2\,\frac{\delta\Phi}{\delta G^{ab}(s,t)},
\end{equation}
self-consistent because $\bm\Sigma$ is itself a functional of the full covariance.
The dressed covariance is thus fixed by self-consistency rather than by a finite
expansion in the strength of the nonlinearity, and it is that preserved
self-consistency that lets a low-order truncation stay accurate when the
nonlinearity is not small. One exact constraint must survive any truncation: the
discounted forward is a martingale, so an admissible closure preserves
$\varphi(-\ii)=1$, imposed as a no-arbitrage condition on the mean equation.

\subsection{The response field and the doubled propagator}
\label{sec:framework-response}

The causal response is not an add-on to the correlation sector: both are
components of a single propagator once the path measure is written in its
Martin--Siggia--Rose--Janssen--De\,Dominicis
(MSRJD)~\cite{MartinSiggiaRose1973,Janssen1976,DeDominicisPeliti1978} form. Enforcing the dynamics
path by path with an auxiliary \emph{response field} $\hat{\bm\phi}$ and averaging
over the noise gives a doubled action whose propagator is a block matrix,
\begin{equation}\label{eq:doubled}
  \mathbf G=
  \begin{pmatrix}\langle\phi\,\phi\rangle & \langle\phi\,\hat\phi\rangle\\[2pt]
  \langle\hat\phi\,\phi\rangle & \langle\hat\phi\,\hat\phi\rangle\end{pmatrix}
  =\begin{pmatrix}C & R\\ R^{\!\top} & 0\end{pmatrix},
\end{equation}
whose diagonal block $C(t,s)=\Cov(\phi_t,\phi_s)$ is the symmetric correlation and
whose mixed block is the retarded \emph{response} $R(t,s)=\delta\bar\phi_t/\delta
h_s$ to a source $h$ added to the drift, causal by construction ($R=0$ for $t<s$);
the response--response block vanishes, the field-theoretic statement of
normalisation and causality. Where $C$ measures fluctuations and prices exotics, $R$
propagates a perturbation forward in time and carries risk. At the Gaussian order
used here the two blocks are corrected by different mechanisms. The response is
$R=(\partial_t-\mathbf J)^{-1}$, in which the drift gradient at the mean,
$\nabla\bm\mu(\bar{\bm\phi})$, is replaced by its Gaussian average
$\mathbf J=\EE_{\mathrm{SCG}}[\nabla\bm\mu]$; this is statistical linearisation, a
moment-closure identity rather than a 2PI self-energy. The genuine 2PI self-energy
acts on the noise instead (the Keldysh block): it replaces the diffusion at the mean
by its Gaussian average $\mathbf D=\EE_{\mathrm{SCG}}[\sigma\sigma^{\!\top}]$, and the
correlation follows as $C=R\,\mathbf D\,R^{\!\top}$. The full doubled construction
is carried out in \supp{S1--S3}.

\subsection{Hartree resummation and where the smile comes from}
\label{sec:framework-hartree}

At Gaussian/Hartree level, mean stationarity retains the local tadpole contraction
while the 2PI functional retains the genuine noise loop (\supp{S1}), making the
fluctuations jointly Gaussian at each time and the self-energy dependent on the
equal-time propagator alone. Its defining feature is that a vertex is evaluated by
the exact Gaussian moment-generating function: a single exponential vertex
$c\,e^{a\phi}$ contributes
\begin{equation}\label{eq:resum}
  \Veff(t)=c\,e^{a\bar\phi}\sum_{n\ge0}\frac{(a^2\gamma)^n}{2^n n!}
  =c\,e^{a\bar\phi+\frac12 a^2\gamma}=c\,\EE[e^{a\phi}],
  \qquad \gamma(t)=G(t,t),
\end{equation}
collapsing the one-loop tadpole and the entire daisy tower into one exponential.
The genuinely non-local skeletons---sunset, basketball---are distinct 2PI graphs
this local resummation does not contain; for exponential vertices they are evaluated
in closed form in \Cref{sec:diagrams}.
A perturbative expansion truncates \eqref{eq:resum} at finite order and ceases to
be accurate once $a^2\gamma=O(1)$; keeping the exponential intact resums those
local Wick contractions exactly, which is what preserves accuracy at long
maturities and large vol-of-vol. When the vertex is a general smooth function the
same expectations are computed by Gauss--Hermite quadrature, converging
exponentially for the smooth integrands that arise here.

A natural objection is that if the fields are Gaussian at each instant the smile
must be flat. It is not, because the Gaussianity lives in the transformed field
coordinates, and non-Gaussianity re-enters through three routes: the nonlinear
back-transformation from fields to prices ($S=e^X$, or the CEV map); mixing, the
average of the conditionally Gaussian log-return over the terminal volatility,
which generates the wings and, with correlation, the skew; and dressing. The
covariance used is not the free (bare) one of the linearised dynamics but the
solution of the Dyson equation \eqref{eq:dyson}, which already contains the
fluctuation corrections generated by the nonlinearity; the effective variance
therefore carries the factor $e^{2\gamma_Y}$, with $\gamma_Y$ the equal-time
log-volatility variance, in full rather than the first few terms of its expansion
that a perturbative method would keep. The first two shape
the conditional law; the third fixes its level at long maturities.

\subsection{The fork: two reductions of the gap equation, and the common recipe}
\label{sec:framework-fork}

Everything downstream is controlled by whether the dressed inverse propagator is
local in time. When the log-volatility process is Markovian, both Gaussian-level
dressings---the averaged drift Jacobian from statistical linearisation, and the noise
self-energy---are diagonal, $\Sigma^{ab}(t,s)=\Sigma_{\text H}^{ab}(t)\,\delta(t-s)$, and the
two-time Dyson equation collapses onto the equal-time covariances
$\gamma_{ab}(t)=G^{ab}(t,t)$, which obey a differential Lyapunov equation whose
coefficients depend on the solution through the Gaussian averages
\begin{equation}\label{eq:riccati}
  \dot{\bm\gamma}=\mathbf J\,\bm\gamma+\bm\gamma\,\mathbf J^{\!\top}+\mathbf D,
  \qquad J_{ab}=\EE_{\text{SCG}}\!\Big[\tfrac{\partial\mu_a}{\partial\phi_b}\Big],
  \quad D_{ab}=\textstyle\sum_c\EE_{\text{SCG}}[\sigma_{ac}\sigma_{bc}],
\end{equation}
coupled to $\dot{\bar\phi}_a=\EE_{\text{SCG}}[\mu_a]$; an $N$-field model needs
$\tfrac12 N(N+3)$ ODEs. When the volatility is rough the kernel has power-law memory,
the two-time covariance it generates is nonstationary with eigenvalues decaying only
as $\lambda_k\sim k^{-(2H+1)}$, and no finite-dimensional Markovian representation is
exact; the framework then retains
the full two-time propagator $G(t,s)$ on a discrete grid. For the rough models of
interest the log-variance field is Gaussian and free, so its propagator is known in
closed form and the work shifts to evaluating the characteristic function as a
Gaussian integral over that field, conditioning on a realisation to render the
log-return conditionally Gaussian (or conditionally exact-CEV in the Lamperti
basis) and doing the outer integral by PCA-ordered quasi-Monte Carlo, or by quadrature
over the two leading modes of the field with the composite pair (integrated variance,
leverage integral) closed by its exact conditional moments. Markovian SABR uses the same
conditional-mode route at $H=\tfrac12$ when the boundary or the correlated drift is
strong.

\begin{figure}[t]
\centering
\includegraphics[width=0.74\textwidth]{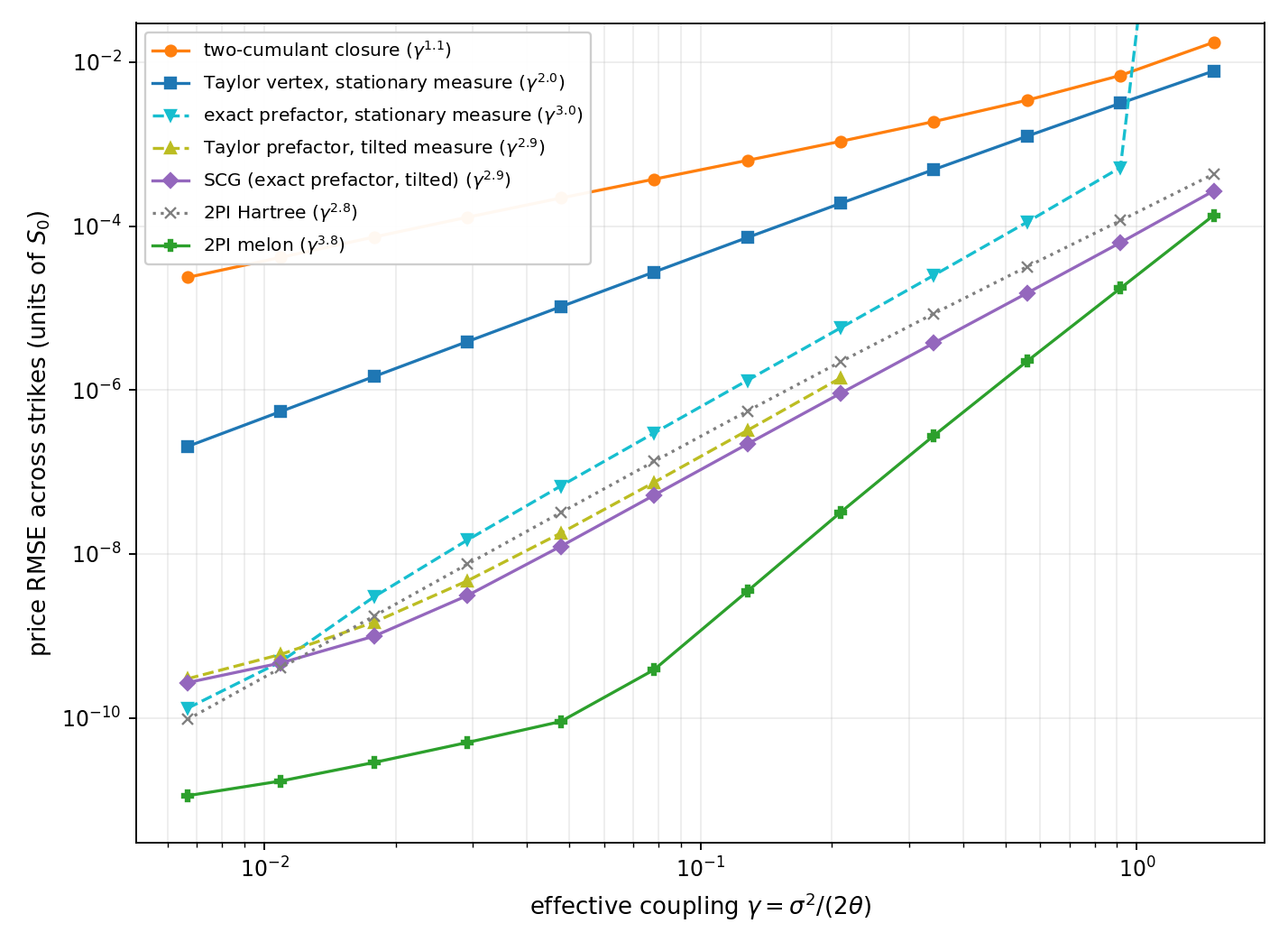}
\caption{Exp-OU error order versus coupling ($\theta=3$, $\mu=-2$, $T=1$,
$\rho=0$): price RMSE against the effective coupling $\gamma=\sigma^2/(2\theta)$, with
fitted slopes on $\gamma\in[0.02,0.6]$. The ablation rows switch the vertex (quadratic
Taylor polynomial, Taylor prefactor, exact prefactor) and the projection measure
(stationary $\mathcal N(0,\gamma)$ or self-consistently tilted $\mathcal N(m,v)$)
separately; the order comes from keeping the vertex exact, the tilt improves the
constant. The 2PI Hartree and melon levels are scored against the tower recursion on
their own 100-step grid, the continuous-time closures against a 400-step tower. At
$\rho\neq0$ the skew vertex enters; the melon level there is reported in
\Cref{tab:ladder}.}
\label{fig:scaling}
\end{figure}

Every application below follows the same five-step recipe on one side of this fork:
identify the Gaussian conditioning field; write the effective action and its
Gaussian-level dressings through \eqref{eq:resum}; solve the gap equation (equal-time ODEs
in the Markovian branch, full two-time propagator in the rough branch); contract to
the target---marginal characteristic function, exotic conditional law, or bucketed
sensitivity, each a contraction of the correlation $C$ and the causal response $R$;
and validate against the strongest independent reference, adding the next skeleton
where a fixed level plateaus. The successive skeletons define a hierarchy---Level~0
the free theory (Black--Scholes), Level~1 the Gaussian closure, Level~2 the melon
skeletons of \Cref{sec:diagrams}. At nonzero correlation in the rough and CEV models the
hierarchy is built instead from the conditional-mode closures, whose truncations---the
number of modes, the moment order of the node law, the number of jumps and of fluctuation
directions---can each be raised. \Cref{fig:scaling} measures the error order on exp-OU
and separates the two ingredients of the Level-1 closure. Keeping the vertex as an
exact moment-generating function rather than its quadratic Taylor polynomial raises
the fitted log--log slope from $2.0$ to $3.0$ under the stationary measure; tilting the
measure self-consistently leaves the order at $\approx2.9$ but lowers the constant by
$4$--$6\times$; with the tilt in place, a second-order Taylor prefactor about the tilted
mean costs only a factor $1.5$ in the constant but breaks down for $\gamma>0.3$. The
evaluated melon level adds about one power, to a slope of $3.8$. These are measured
orders at one parameter point, not theorems.

\section{The Markovian Branch: exp-OU, SABR and Stochastic Rates}
\label{sec:markovian}

In the Markovian branch the self-energy is local and the gap equation is the
equal-time system \eqref{eq:riccati}: a handful of ODEs per smile. We take the two
principal non-affine models---exp-OU (lognormal) and SABR (CEV)---and then show
the same closure carrying several factors in an FX local-volatility spot with two
stochastic rate curves.

\subsection{Exponential-OU}
\label{sec:expou}
The exp-OU model is the Markovian lognormal proving ground: the log-volatility is a
stationary Ornstein--Uhlenbeck process $X$ and the variance is its exponential,
$v=e^{X}$, driving $\dd S/S=\sqrt v\,\dd W^S$ with
$\langle\dd W^S,\dd W^X\rangle=\rho\,\dd t$. The exponential vertex has the
universal coupling $e^\mu$ at every order, so the tower must be resummed.
Evaluating the exponential vertex through \eqref{eq:resum} gives the local dressing
$\Sigma_{\text H}=e^{2\mu+G^X(0)}$, which is
frequency-independent and therefore a pure mass renormalisation: the dressed
propagator keeps the OU form with a shifted pole $\theta\to\thetaeff$,
$\thetaeff^2=\theta^2+2\theta\Sigma_{\text H}/\sigma^2$, reached by damped iteration
of a scalar map in a few tens of steps. For pricing, the $u$-dependent conditional
wave function satisfies an imaginary-time Schrödinger equation with an exponential
potential (the integrated-variance mass term) and a correlation advection (the
skew vertex); the SCG truncation adopts a log-quadratic ansatz
$\psi=\exp(A+Bz+\tfrac12 Cz^2)$ and projects onto $\{1,z,z^2\}$ under the effective
Gaussian measure, closing to three ODEs in $(A,B,C)$ with the variance vertex
resummed exactly by $\EE[e^{az}]=e^{am+a^2v/2}$,
\begin{equation}\label{eq:expou-scg}
\begin{aligned}
  C' &= \sigma^2 C^2 - 2\theta C - w\Veff
       + \lambda V_{1/2}\bigl(C + \tfrac{1}{4}P\bigr), \\
  B' &= -(\theta - \sigma^2 C)B - w\Veff(1-m)
       + \lambda V_{1/2}\bigl(C(1-m) + \tfrac{1}{4}P(2-m)\bigr), \\
  A' &= \tfrac{\sigma^2}{2}(C + B^2) - w\Veff\bigl(1 - m + \tfrac{m^2-v}{2}\bigr)
       + \lambda V_{1/2}\bigl(P\bigl(1-\tfrac m2+\tfrac{m^2}{8}\bigr)
         + C\bigl(\tfrac{m^2}{2}-m\bigr)\bigr),
\end{aligned}
\end{equation}
with $w=\tfrac12(\ii u+u^2)$, $\lambda=\ii u\rho\sigma$, $P=B+Cm$,
$\Veff=e^{\mu+m+v/2}$, $V_{1/2}=e^{\mu/2+m/2+v/8}$, and effective mean and variance
$m=B\gamma/(1-\gamma C)$, $v=\gamma/(1-\gamma C)$ (the skew vertex $V_{1/2}$ vanishes
at $\rho=0$). The characteristic function is the Gaussian integral
$\varphi(u;T)=e^{A+B^2\gamma/(2(1-\gamma C))}/\sqrt{1-\gamma C}$, inverted to prices
by the Fang--Oosterlee COS method~\cite{FangOosterlee2008}; the full projection is
given in \supp{S4}.

\begin{figure}[t]
\centering
\includegraphics[width=\textwidth]{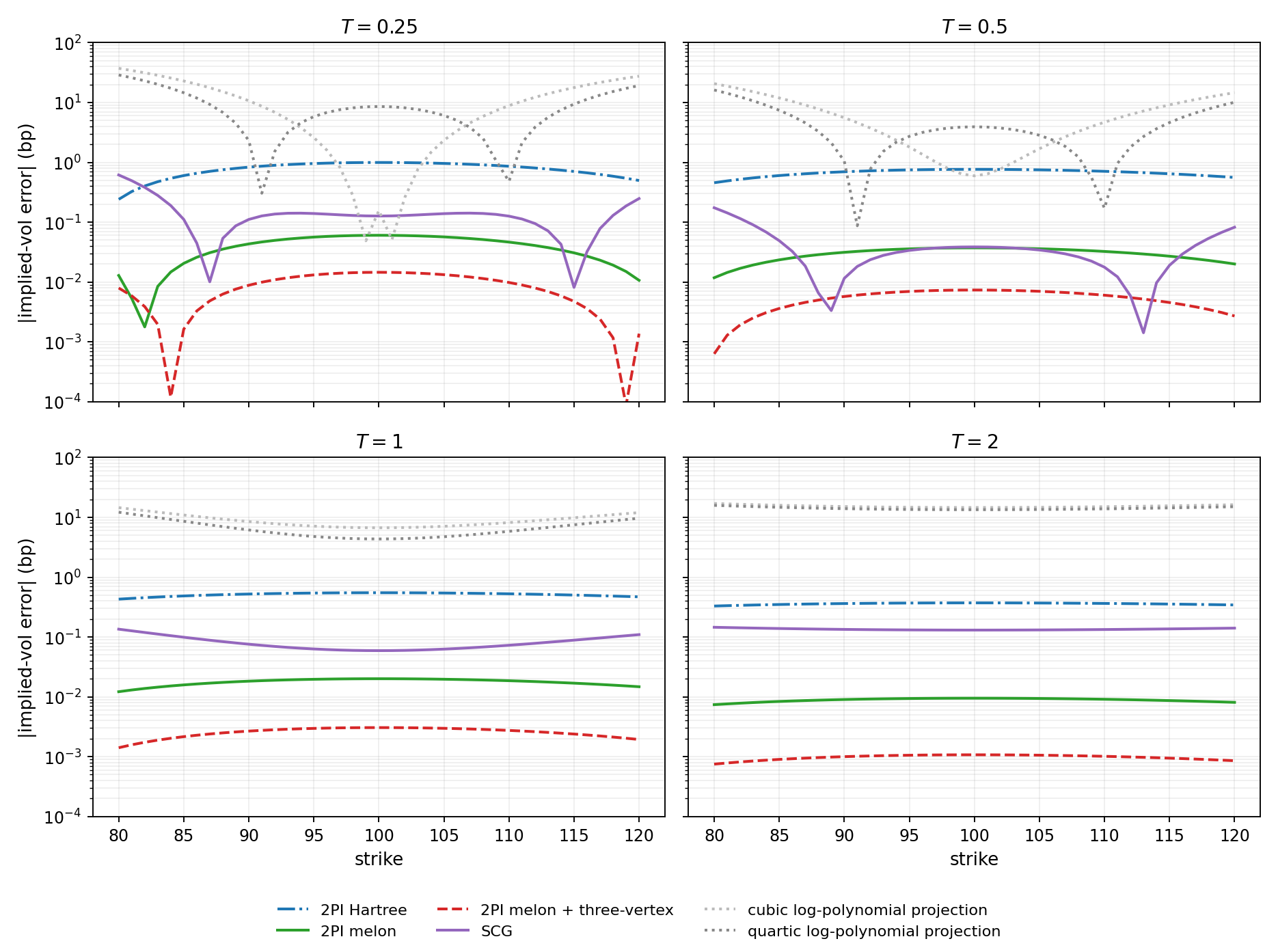}
\caption{Exp-OU ($\theta=3$, $\sigma=1.5$, $\mu=-2$, $\rho=0$): absolute implied-vol
error of each level against the exact tower recursion. The evaluated 2PI skeletons
converge level by level---Hartree $0.36$--$0.81$\,bp, melon $0.01$--$0.04$\,bp,
melon plus three-vertex skeletons below $0.01$\,bp---while the SCG sits at
$0.05$--$0.19$\,bp and the cubic and quartic log-polynomial projections at
$6$--$16$\,bp. The 2PI levels are scored on their own $100$-step-per-year grid; the
SCG and the projections against an eight-times finer tower, whose own time-step error
is below $0.1$\,bp.}
\label{fig:expou}
\end{figure}

Against a tower PDE and a $2^{20}$-path Sobol reference the SCG achieves
$0.1$--$0.2$\,bp RMSE across all strikes and maturities at $\rho=0$, and
sub-basis-point accuracy at $\rho\in\{-0.7,-0.4,0.4\}$ with the skew shape faithfully
reproduced, at one three-ODE solve per smile (well under a millisecond in the single-core Python reference\footnote{\timingnote}). Exp-OU is
therefore essentially solved at Level~1 in this cell, and the evaluated skeleton
ladder of \Cref{sec:diagrams} converges onto the exact characteristic function level
by level (\Cref{fig:expou}): the 2PI Hartree level is within $0.4$--$0.8$\,bp, the melon
level within $0.01$--$0.04$\,bp and the three-vertex level within $0.01$\,bp, each
level with the opposite sign to the one before. Cubic and quartic terms added to
the log-polynomial ansatz for $\psi$, by contrast, do not improve on Level~1 here
($6$--$16$\,bp): they are projection orders, not skeletons, and we no longer use them.

\subsection{SABR}
\label{sec:sabr}
SABR adds the CEV nonlinearity $F^\beta$, whose back-transform carries the skew, and
a free volatility field, $\dd F=\alpha F^\beta\,\dd W^F$,
$\dd\alpha=\nu\alpha\,\dd W^\alpha$, $\langle\dd W^F,\dd W^\alpha\rangle=\rho\,\dd t$.
It is the model on which the framework works hardest. The log-volatility enters the
action only through its couplings to the forward, so no 1PI diagram dresses the
$\alpha$-propagator: $\Sigma^\alpha=0$ and $G^\alpha(t,s)=\nu^2\min(t,s)$ to all
orders. The terminal volatility therefore has an exact lognormal marginal and can be
integrated out by an outer Gauss--Hermite quadrature, conditional on which the
process is an exact Brownian bridge. For $\beta<1$ the Lamperti transform
$Z=F^{1-\beta}/(1-\beta)$ concentrates all CEV nonlinearity in a Bessel drift and
leaves the pure stochastic-volatility diffusion, making $Z$ the natural SCG
coordinate.

A purely Gaussian inner closure prices each node from a Gaussian in $Z$-space but
strips the CEV skew, degrading from $\sim$2\,bp to $\sim$125\,bp at $\beta=0.3$. The
remedy keeps the Level-1 volatility statistics but prices the forward from the
\emph{exact} CEV transition density (a noncentral-$\chi^2$ survival). Conditional on
the volatility path, the part of the forward's noise correlated with the volatility
is known and enters the Lamperti coordinate as a time-dependent displacement.
Approximating that displacement by an effective start,
$Z_0^{\text{eff}}=Z_0+(\rho/\nu)(\alpha_T-\alpha_0)$ (Islah~\cite{Islah2009}), leaves
a Bessel process in the clock $\tau=(1-\rho^2)\int_0^T\alpha_t^2\,\dd t$ whose drift
coefficient is correlation-enhanced,
$\beta'=2c'/(1+2c')>\beta$ with $c'=c/(1-\rho^2)$ and $c=\beta/2(1-\beta)$, and the
forward is priced from the exact CEV$(\beta')$ Green's function. The transition
density is exact for this auxiliary CEV model; the reduction from correlated SABR to
it is not. Using
$\beta'$ repairs the $O(\rho^2)$ exponent defect of a naive constant-$\beta$ time
change while preserving the exact $\rho=0$ limit, and the martingale (Ward) identity
$\EE[F_T]=F_0$ is restored by a constant shift of the per-node Lamperti starts. Two
corrections handle the residual corners: a displacement \emph{booster} (a commutator
shift between the continuous displacement ramp and the endpoint displacement) in the
positive-$\rho$, high-$\beta$ corner, and a boundary-proximity (Stokes-inspired)
indicator---one auxiliary ODE per outer node---that routes the negative-$\rho$,
low-$\beta$, long-maturity cells to a two-dimensional multi-slice lattice. The full
pricing stack, survival function and routing are given in \supp{S5}.

The reference absorbs the forward at zero with a Brownian-bridge survival weight,
$w\leftarrow w\bigl(1-\exp(-2F_kF_{k+1}/(\alpha_k^2F_k^{2\beta}\Delta t))\bigr)$, rather
than by clamping. Clamping misses crossings within a step, an $O(\sqrt{\Delta t})$ error
that step extrapolation does not remove; at $\rho=0$, where the conditional law is
exactly CEV and an exact reference is available (the Schr\"oder price averaged over
exact lognormal volatility paths), a clamped scheme is $4$\,bp off at $\beta=0.1$,
$T=5$ and $10$\,bp at $T=10$, while the bridge-weighted scheme agrees with the exact
reference to $0.14$--$0.98$\,bp in all twelve $\rho=0$ cells. With $2^{17}$ paths,
$16$ scrambles, a martingale control variate and $512$ steps the standard error is
$0.13$--$0.62$\,bp, and the change from $256$ to $512$ steps is below $2$\,bp except in
three cells at $T=10$ with $\beta\ge0.5$ ($2.5$--$5.3$\,bp).

\begin{figure}[t]
\centering
\includegraphics[width=\textwidth]{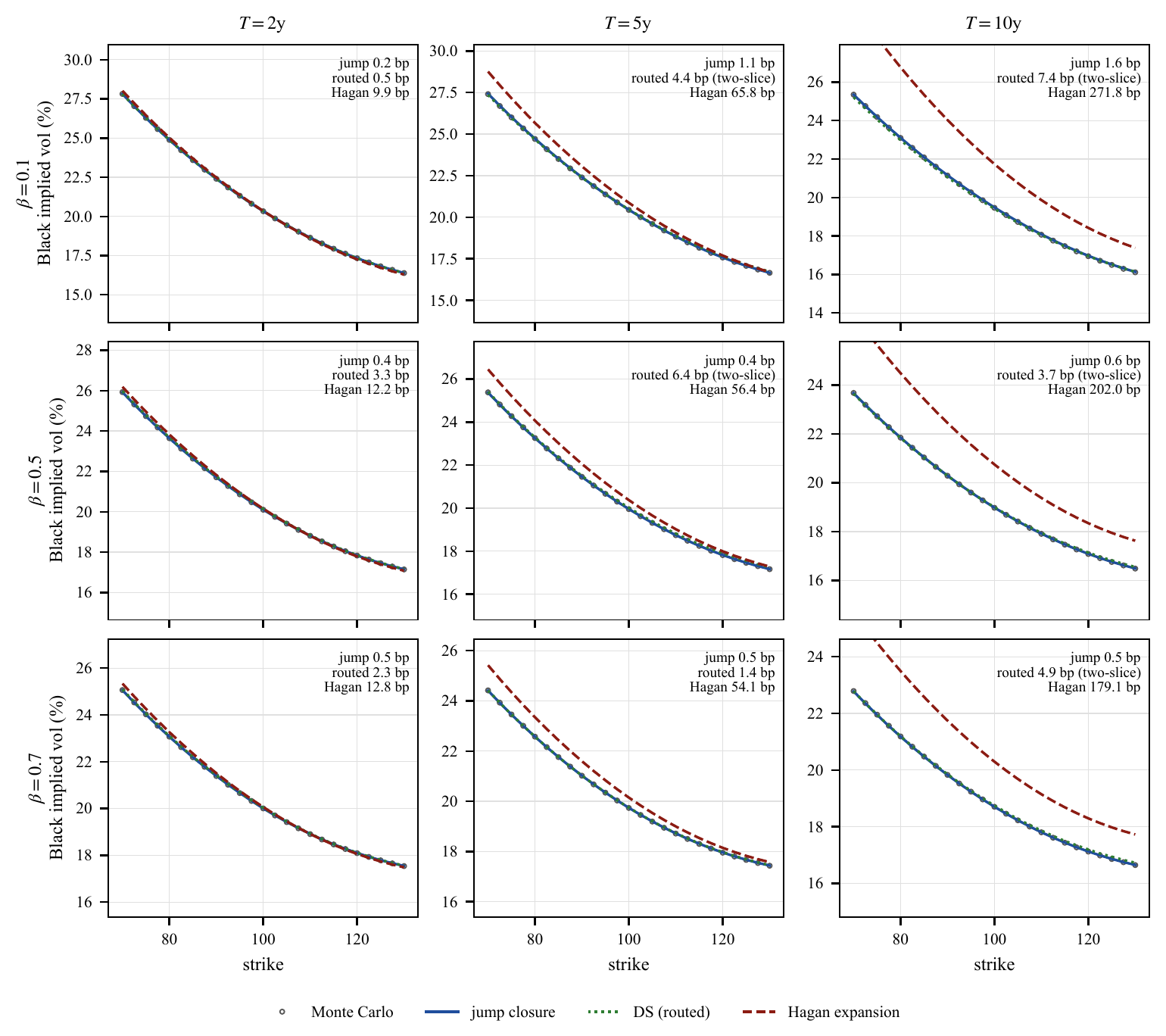}
\caption{SABR smiles ($F_0=100$, $\alpha_0=0.2F_0^{1-\beta}$, $\nu=0.4$, $\rho=-0.5$) for
$\beta\in\{0.1,0.5,0.7\}$ (rows) and $T\in\{2,5,10\}$\,y (columns). Markers and band:
bridge-absorbed Sobol reference ($16$ scrambles of $2^{17}$ paths, $512$ steps) with $\pm2$
standard errors. Solid: jump closure. Dotted: routed stack (``two-slice'' marks cells
routed to the multi-slice lattice). Dashed: Hagan's expansion, which leaves the panel at
$T=10$.}
\label{fig:sabr}
\end{figure}

Over $\beta\in\{0.1,0.3,0.5,0.7\}$, $\rho\in\{-0.9,\dots,0.5\}$, $T\in\{2,5,10\}$, the
boundary-proximity routing recovers most of the way to a per-cell oracle at no extra
cost (\Cref{tab:sabr-routing}), for a grid-mean $7.1$\,bp against $74.9$\,bp for Hagan's
expansion, and is more accurate than Hagan in $69$ of the $72$ cells (the exceptions are
at $T=2$, $\rho=-0.9$). At $T=2$ the routed error is $0.5$--$8.5$\,bp; at $T=5$ it is
$0.7$--$29$\,bp against $18$--$66$\,bp for Hagan (\Cref{fig:sabr}). The worst cell
($\beta=0.1$, $\rho=-0.9$, $T=10$, $64$\,bp) is the boundary corner, where the forward
reaches zero early on most volatility paths. The first version of this article attributed
that error to the model; the jump closure below prices the cell to $1.3$\,bp, so it is a
limitation of the routed stack. The routing constants were chosen on this grid, so these figures are
in-sample. On a held-out grid of $24$ cells lying between the tuning values,
$\beta\in\{0.2,0.4,0.6\}$, $\rho\in\{-0.8,-0.4,-0.1,0.3\}$, $T\in\{3,7\}$, with no
constant retuned, the routed pricer has mean RMSE $4.3$\,bp (maximum $16.6$\,bp)
against $53.5$\,bp (maximum $133$\,bp) for Hagan, and is more accurate than Hagan in
every cell. At $T=3$ it ranges over $0.4$--$5.3$\,bp and at $T=7$ over $1.0$--$16.6$\,bp.
Choosing the better of the two pricers per cell would give a mean of $4.0$\,bp; the
largest routing losses are at $\rho=0.3$, $T=7$, where the dressed pricer is chosen and
the moment-matched alternative is about $4$\,bp closer.

\begin{table}[t]
\centering
\begin{tabular}{@{}lrr@{}}
\toprule
\textbf{Routing policy} & \textbf{Grid-max (bp)} & \textbf{Grid-mean (bp)} \\
\midrule
All-dressed (single slice) & 316.2 & 27.1 \\
Boundary-proximity routing & 63.5 & 7.1 \\
Oracle (best method per cell) & 63.5 & 6.2 \\
Hagan lognormal expansion & 288.5 & 74.9 \\
\bottomrule
\end{tabular}
\caption{SABR implied-vol RMSE (25 strikes, $70$--$130$) over 72 cells against the
bridge-absorbed Monte-Carlo reference. Boundary-proximity routing matches the per-cell
oracle in grid-max and, in grid-mean, closes $96\%$ of the gap between the single-slice
engine and the oracle, at no extra pricing cost.}
\label{tab:sabr-routing}
\end{table}

\paragraph{The jump closure.}
The routed stack applies the whole correlated displacement at the start. Conditional on
the volatility path the Lamperti variable solves $\dd Z=\dd B-c'Z^{-1}\dd\tau+\dd D$ on the
clock $\tau$, killed at zero, where $D_\tau$ is the displacement
$\rho\int\alpha\,\dd W^\alpha$ accumulated up to clock $\tau$. The exact conditional value
$U$ therefore obeys a Dyson (Duhamel) equation around the absorbed-CEV propagator $P_0$ of
the displaced-start problem, $U=U_0+\int_0^\tau P_0\bigl(\dot r\,\partial_zU\bigr)\dd s$,
with $r(s)$ the displacement still to come at clock $s$; the Islah start is its zeroth
order, and its first-order term is the remaining displacement integrated against the
gradient of the absorbed value function. The jump closure resums this series by
composition: it applies $D$ in jumps $W_j=\sum_k\Delta D_k\,\ell_j(S_k)$ at the
Chebyshev--Lobatto nodes $s_j$ of the clock fraction, the projection of the displacement
increments on the Lagrange basis $\ell_j$, and propagates the exact absorbed-CEV law between
jumps, killing any path that a jump takes below zero. The $W_j$ are the first-order Duhamel
weights, so the first-order term is reproduced and the higher orders are carried by the
composition of exact kernels. The volatility field enters through two Gaussian modes, the
linearisations of $\log\int\alpha^2\dd t$ and of $\int\alpha\,\dd W^\alpha$; at each mode node a
moment-matched law for $(\int\alpha^2\dd t,D_T)$ supplies inner points, the jumps at each
point are their conditional means given the node, the point and $D_T$, and their residual
covariance is integrated along its leading eigen-directions by a rule with non-negative
weights and exact variances. The jump kernel is smooth in the inner variables, so the
price is the displaced-start mixture on the full inner rule plus the jump-minus-start
difference on a coarse rule, with a constant shift of the starts restoring
$\EE[F_T]=F_0$. Seven Lobatto jumps suffice below ten years; at $T=10$ we use $17$ for
$\rho\le0$ and, for $\rho>0$, $13$ jumps or $32$ piecewise-linear hats at $\beta\le0.5$.
The construction and its settings are given in \supp{S11}.

\begin{table}[t]
\centering
\small
\begin{tabular}{@{}lrrrrrr@{}}
\toprule
 & \multicolumn{2}{c}{\textbf{routed stack}} & \multicolumn{2}{c}{\textbf{jump closure}} & \multicolumn{2}{c}{\textbf{Hagan}} \\
\cmidrule(lr){2-3}\cmidrule(lr){4-5}\cmidrule(lr){6-7}
$T$ & mean & max & mean & max & mean & max \\
\midrule
2   & 2.8 & 8.5 & 0.42 & 0.72 & 8.8 & 12.8 \\
5   & 5.4 & 29.3 & 1.22 & 6.54 & 43.6 & 65.8 \\
10  & 13.0 & 63.5 & 1.28 & 4.36 & 172.1 & 288.5 \\
all & 7.1 & 63.5 & 0.97 & 6.54 & 74.9 & 288.5 \\
\midrule
held-out & 4.3 & 16.6 & 1.06 & 5.18 & 53.5 & 133 \\
\bottomrule
\end{tabular}
\caption{SABR implied-vol RMSE (bp, 25 strikes $70$--$130$) against the bridge-absorbed
Monte-Carlo reference: the $72$-cell grid of \Cref{tab:sabr-routing} by maturity, and the
$24$ held-out cells ($\beta\in\{0.2,0.4,0.6\}$, $\rho\in\{-0.8,-0.4,-0.1,0.3\}$,
$T\in\{3,7\}$) on which no setting was chosen.}
\label{tab:sabr-jump}
\end{table}

\Cref{tab:sabr-jump} shows the jump closure below $2$\,bp in $67$ of the $72$ cells and
below $5$\,bp in $71$, more accurate than Hagan in every cell, and as accurate on the
held-out cells as on the grid. It removes the three residual sources of the routed
stack: the $\rho=0$, $T=10$ cells fall from $7$--$8$ to $0.5$--$2.2$\,bp, the
$\rho=0.5$, $T=10$ cells from $13$--$30$ to $0.2$--$1.5$\,bp, and the boundary corner from
$64$ to $1.3$\,bp. Its worst cells are at $\beta\le0.3$, $\rho\le-0.7$ ($6.5$\,bp at
$\beta=0.1$, $\rho=-0.9$, $T=5$), and most of its errors are of the size of the reference
standard error. The routed stack prices from non-negative mixture weights; the jump
closure's assembly is a signed control variate, and its positivity is checked rather than
structural: over all cells the smallest butterfly is $-9\times10^{-13}$, rounding level.
The cost is $8.5$\,s per smile with seven jumps, $13.7$\,s with $17$ and $26.7$\,s with $32$
hats, on one thread.\footnote{\timingnote}

\begin{figure}[t]
\centering
\includegraphics[width=\textwidth]{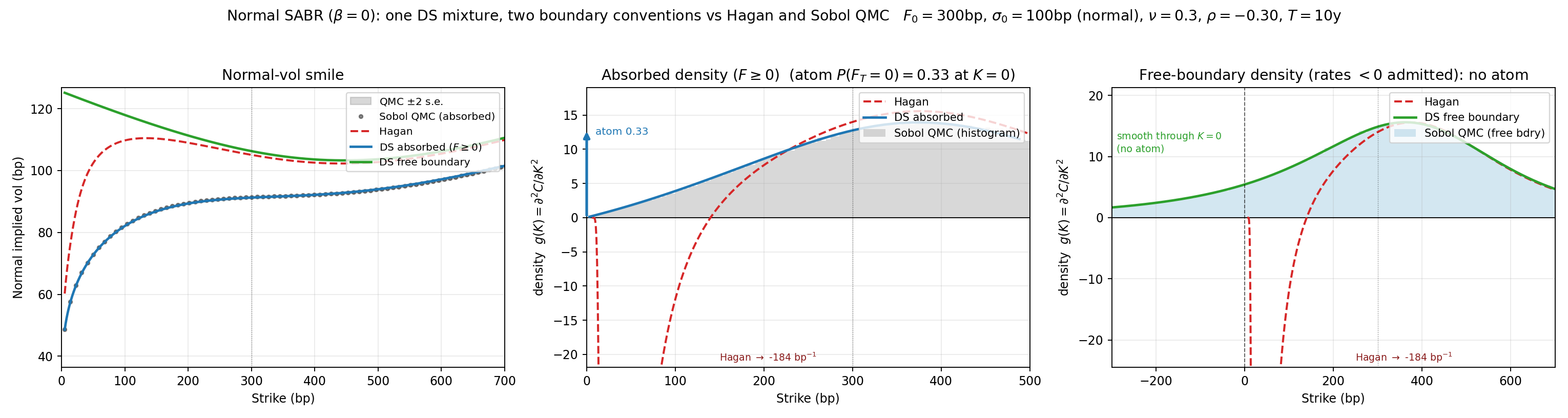}
\caption{Normal SABR ($\beta=0$, absorbed non-negative-rates convention) in a rates
framework with a realistic skew. \emph{Left:} the DS multi-slice pricer (blue) sits on
the bridge-absorbed Sobol reference to within $0.4$\,bp while Hagan overstates the
low-strike wing. \emph{Right:} the Breeden--Litzenberger density---DS non-negative by
construction and matching the QMC histogram, Hagan plunging to $\approx-184$ (a
butterfly arbitrage). The Hagan curve is the free-boundary asymptotic expansion, not
the absorbed transition law, so its negative wing is that expansion's small-strike
breakdown rather than a same-law comparison.}
\label{fig:normalsabr}
\end{figure}

\paragraph{Arbitrage-free densities in the normal case.}
Normal SABR admits two conventions for what happens when the forward reaches zero,
and we are explicit about which the figures use. Under the standard
\emph{free-boundary} convention the forward simply crosses zero: conditional on the
volatility path its law is Gaussian on the whole real line, negative rates are
allowed, and the terminal distribution has a continuous density with no special
behaviour at $F=0$. Under the \emph{absorbed} (non-negative rates) convention a path
that reaches $F=0$ stays there until maturity, so the terminal law splits into a
continuous density on $F>0$ plus a point mass (an \emph{atom}) at $F=0$ equal to the
probability of having been absorbed. \Cref{fig:normalsabr} adopts the absorbed
convention. There, at $\beta=0$ ($\sigma_0$ the normal vol) the DS construction
cannot produce a negative density, for a structural reason rather than a repair: it
prices every terminal-volatility node from the exact absorbed-CEV transition law---a
genuine probability density---and mixes those laws with non-negative Gauss--Hermite
weights, so the Breeden--Litzenberger density $g(K)=\partial_K^2 C\ge0$ holds node
by node. On a low-rates cell with realistic skew ($F_0=300$\,bp, $\sigma_0=100$\,bp
normal, $\nu=0.3$, $\rho=-0.3$, $T=10$\,y) the DS multi-slice pricer sits on a
bridge-absorbed Sobol reference to within $0.4$\,bp, reproduces the genuine atom
$\PP(F_T=0)\approx0.33$, and stays non-negative everywhere, whereas Hagan dips to
$g\approx-184$ over the $6$--$139$\,bp wing---a static butterfly arbitrage
(\Cref{fig:normalsabr}). This is a structural, not a same-transition-law, comparison:
Hagan is the free-boundary asymptotic expansion, so its negative wing is the
well-known small-strike breakdown of that expansion rather than a mispriced version of
the absorbed law; the point is that the absorbed DS law is arbitrage-free by
construction exactly where the free-boundary expansion is not.

\subsection{FX local volatility with two stochastic curves}
\label{sec:fxhw}
Nothing restricts the framework to a single stochastic-volatility factor. A Dupire
local-volatility FX spot with a domestic and a foreign Hull--White short
rate~\cite{Dupire1994,HullWhite1990,Piterbarg2006}---three correlated drivers---stays
within the same equal-time closure. The two rates are treated as \emph{free fields}:
their Hull--White second moments enter the spot equation as data, and only the spot
carries interaction through $\sigma_{\mathrm{loc}}(t,S_t)$. For the domestic rate this
is exact. The foreign rate carries a quanto drift that depends on the spot, so taking
its variance and its covariance with the domestic rate from the free formulas is an
additional approximation, exact when $\rho_{Sf}=0$. The SCG
closure keeps $(X,r_d,r_f)$ jointly Gaussian, carrying the three means, the spot
variance, the two cross-covariances, and the third and fourth spot cumulants to
dress the characteristic function beyond Gaussian; every local-volatility moment is a
one-dimensional Gauss--Hermite quadrature against the current Gaussian, and pricing
is under the $T$-forward measure. The whole smile costs eight ODEs integrated once;
the derivation and the regularised cumulant Fourier kernel are given in \supp{S6}.

\begin{figure}[t]
\centering
\includegraphics[width=\textwidth]{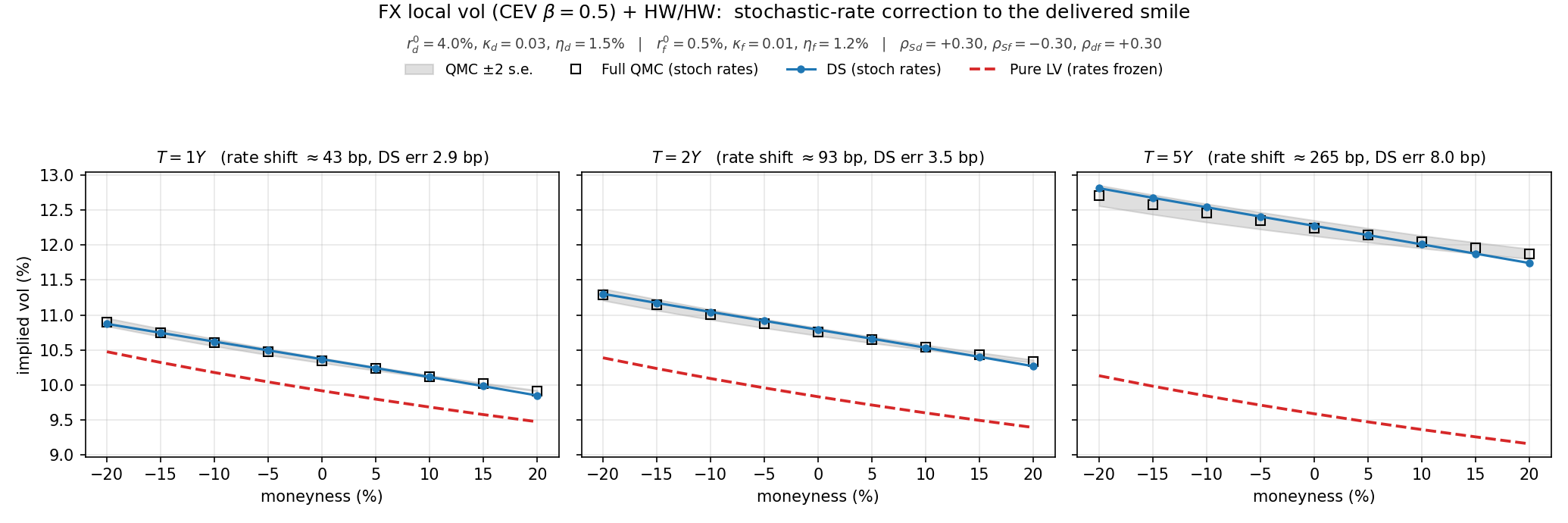}
\caption{Delivered terminal smiles for an FX local-volatility model (CEV,
$\beta=0.5$) with two correlated Hull--White curves at $1$, $2$, $5$\,y. The DS/SCG
engine (blue) sits on the full stochastic-rate Sobol QMC (open squares) within a few
bp, while the identical local volatility with \emph{frozen} rates (red dashed) sits
far below: the gap is the stochastic-rate correction, $43$/$93$/$265$\,bp at
$1$/$2$/$5$\,y.}
\label{fig:fxhw}
\end{figure}

The stochastic-rate correction is large and grows with maturity---about $43$\,bp at
one year, $93$\,bp at two, $265$\,bp at five---so a local-vol model calibrated with
deterministic discounting misprices the wing by that much. The SCG engine tracks the
full stochastic-rate Sobol smile to within $2.9$, $3.5$ and $8.0$\,bp RMSE across
the three maturities, capturing essentially all of the correction the frozen-rate
model misses, at eight ODEs against seconds of simulation per tenor (\Cref{fig:fxhw}).
The same conditional moments deliver the extended-Dupire conditional drift at no extra cost;
the leverage calibration built on it is left to future work.

\section{The Rough Branch: Two-Time Propagator}
\label{sec:rough}

Rough volatility replaces the OU or GBM log-variance with a Volterra Gaussian
process of power-law kernel $K(\tau)\propto\tau^{H-1/2}$ and Hurst
$H\approx0.1$~\cite{GatheralJaissonRosenbaum2018}. The line runs from the long-memory
continuous-time models of Comte and Renault~\cite{ComteRenault1998}, through the
short-time martingale expansions of Al\`os--Le\'on--Vives and
Fukasawa~\cite{AlosLeonVives2007,Fukasawa2011}, to the rough-Bergomi pricing
programme~\cite{BayerFrizGatheral2016} and its efficient simulation by hybrid and
turbocharged Monte Carlo~\cite{BennedsenLundePakkanen2017,McCrickerdPakkanen2018}.
The kernel
has power-law memory, the resulting two-time covariance is nonstationary, and no
finite-dimensional Markovian representation is exact, so the ODE reduction fails and the framework retains the full two-time propagator on
a grid. The free Gaussian field makes this tractable: conditioning on a realisation
renders the log-return conditionally Gaussian (or conditionally exact-CEV), and the
outer field integral is done by PCA-ordered quasi-Monte Carlo or, deterministically, by
quadrature over the field's leading modes with a moment closure for the rest. Only the
volatility-field sampler is model-specific; everything else is shared with the
Markovian branch.

\subsection{Rough Heston as an exactly-transformable control}
\label{sec:rheston}
Rough Heston is affine, so its characteristic function follows from a fractional
Riccati equation~\cite{ElEuchRosenbaum2019}, which is already exact, so no closure is
involved. It serves only as a check on the two-time numerics: the non-local fractional
Riccati equation solved on the same grid by the fixed-point (Picard/Newton) scheme used
elsewhere, with its $O(\Delta t^{H+1/2})$ discretisation, reproduces option
prices/implied volatilities obtained from the transform to within $1$--$5$\,bp. The
$O(\Delta t^{H+1/2})$ rate is that of the rectangle rule for the fractional integral
with the singular kernel $\tau^{H-1/2}$, not a result of this paper.

\subsection{Rough Bergomi}
\label{sec:rbergomi}
Rough Bergomi is the central rough application:
$v_t=v_0\exp(\eta\widehat W_t-\tfrac12\eta^2 t^{2H})$ with $\widehat W$ the Gaussian
Volterra field, so the log-variance $Y_t=\eta\widehat W_t-\tfrac12\eta^2 t^{2H}$ is a
Gaussian field with deterministic mean $-\tfrac12\eta^2 t^{2H}$ and a known two-time
covariance $G(t,s)$---its random component $\eta\widehat W_t$ is centred---and needs no dressing.
Splitting $W^S=\rho W^v+\sqrt{1-\rho^2}W^\perp$ decomposes the log-return into
channels that are conditionally deterministic in their integrands given the path $Y$,
\begin{equation}\label{eq:rB-decomp}
\begin{gathered}
  X_T = -\tfrac12 V + \rho J + \sqrt{1-\rho^2}\,K,
  \qquad
  V = v_0\!\int_0^T\! e^{Y_t}\dd t,\\
  J = \sqrt{v_0}\!\int_0^T\! e^{Y_t/2}\dd W_t^v,
  \qquad
  K = \sqrt{v_0}\!\int_0^T\! e^{Y_t/2}\dd W_t^\perp,
\end{gathered}
\end{equation}
so conditioning on the entire volatility path $W^v$ (equivalently the increments that
generate the discretised field $Y$), both $V$ and the correlated integral $J$ are known
pathwise---$J$ is a nonlinear, non-Gaussian functional of $W^v$, but fixed once the path
is---while $W^\perp$ is independent. The return is therefore \emph{exactly} Gaussian
conditional on the path, $X_T\mid W^v\sim\mathcal N(-\tfrac12 V+\rho J,\,(1-\rho^2)V)$ (the
Romano--Touzi mixing identity~\cite{RomanoTouzi1997}), and
$\varphi(u)=\EE[\exp(\ii u(-\tfrac12 V+\rho J)-\tfrac12 u^2(1-\rho^2)V)]$ is exact. The DS
conditional-RQMC evaluation takes this field average over the discrete field, sampled in
its principal (Karhunen--Lo\`eve) coordinates $\xi$ ordered by decreasing eigenvalue; the
unresolved within-step increments are carried by an \emph{approximate} conditional-Gaussian
residual---Gaussian once the integrand $e^{Y_t/2}$ is frozen on the resolved field---giving
$X_T\mid\xi\sim\mathcal N(m,s^2)$ with
$m=-\tfrac12 V+\rho\mu_J$ and $s^2=\rho^2\sigma_J^2+(1-\rho^2)V$, which recovers the pathwise
identity as the grid refines ($\sigma_J^2\to0$ as $\Delta t\to0$). Here $V$ is a nonlinear functional of the
resolved field realisation, while $\mu_J$ and $\sigma_J^2$ are path-dependent contractions of the
conditional mean and covariance of the unresolved Brownian increments; that conditional law is
fixed by the joint covariance of the leverage driver increments and the discretised field, not
by the field covariance alone (explicit conditioning in \supp{S6}). Starting from this exact pathwise identity, the implementation introduces
separately controlled numerical approximations: time discretisation of the field (at
$O(\Delta t^{H+1/2})$), the frozen-integrand conditional-Gaussian residual, and the outer
RQMC with COS inversion downstream; the full discrete field is retained at rank $N$, the
PCA ordering serving only to concentrate the RQMC effective dimension, so no rank-truncation
error enters. The outer integral over the field is done by
PCA-ordered Sobol; because the engine runs on the native two-time covariance, one
field ensemble is amortised across strikes and, later, barriers and reset dates.
This engine evaluates the field average by conditioning rather than by closure: the field
is free, the conditional law is exact up to the within-step residual, and what the
framework supplies is the two-time covariance and its reuse; the composite closure below
replaces the outer sampling by quadrature over the leading modes. \Cref{tab:rbergomi} places the accuracy against Sobol Monte
Carlo on the same time grid: the PCA-resolved conditional RQMC holds to $\sim$$1$--$3$\,bp
across mild to extreme coupling, while a two-cumulant Gaussian characteristic function
for the log-return and a four-cumulant NIG kernel carry errors of tens to hundreds of
basis points. The first of these is not the 2PI Hartree level: it matches only the
first two cumulants of $X_T$ and so has almost no smile, which is why it misses by
$186$\,bp even in the mild regime.

\paragraph{The composite closure.}
The characteristic function needs only the joint law of the two scalars $(V,J)$, since
$K\mid(V,J)\sim\mathcal N(0,V)$, and that law can be built without sampling the field.
Write $\phi_i=Y_{t_{i-1}}$ and $w_i=\Delta W^v_i$, jointly Gaussian with covariance
$\Sigma$. Two linear functionals carry most of the variation of $(V,J)$:
$u_1\propto\sum_k\EE[v_{t_k}]\,\phi_k$, the linearisation of $\log V$, and
$u_2\propto\sum_k\EE[\sqrt{v_{t_k}}]\,w_k$, the linear part of $J$, orthonormalised in
$\Sigma$. Given $u$ the field is still Gaussian, with mean linear in $u$ and covariance
$\Sigma-R^\top R$, $R=\Cov(u,(\phi,w))$, so the nine moments $\EE[V^n]$ ($n\le3$),
$\EE[J]$, $\EE[JV]$, $\EE[J^2]$, $\EE[JV^2]$, $\EE[J^2V]$ and $\EE[J^3]$ given $u$ are finite
sums of lognormal expectations, evaluated as matrix products at $O(N^3)$ cost. At each
node of a $16\times16$ Gauss--Hermite rule in $u$ a nine-parameter law is fitted to them,
\begin{equation}\label{eq:composite}
  \log V=\mu+s\,\frac{g+\lambda(g^2-1)}{\sqrt{1+2\lambda^2}},\qquad
  J=\sqrt V\Bigl(a(g)+e^{\ell(g)/2}\,\frac{h+\kappa(h^2-1)}{\sqrt{1+2\kappa^2}}\Bigr),
\end{equation}
with $g,h$ independent standard normals, $a$ quadratic and $\ell$ linear in $g$. The
characteristic function becomes a finite mixture of Gaussian transforms and the vanilla a
mixture of Black prices, with the forward normalised once after the mode integral. In
field-theory terms the closure works with the composite operators $(V,J)$ rather than
with the field, in the spirit of the effective action for composite
operators~\cite{CJT1974}, and treats their first three moments exactly given the two
leading modes of the free field. Beyond the time grid its only approximations are
the moment-matched node law and the mode rule, both of which can be raised; it is
deterministic and smooth in the parameters, and a seven-strike smile costs about $1$\,s on
one thread with $N=200$ (\supp{S11}).

Against the $2^{24}$-path conditional reference of \Cref{tab:rbergomi}, whose standard
error is at most $0.12$\,bp, the composite closure is approximately at the reference
resolution in each regime (RMSE $0.002$, $0.058$ and $0.115$\,bp against standard errors of
$0.013$, $0.04$ and $0.10$\,bp), where the 16k-node
RQMC average is at $0.2$--$2.9$\,bp. Over a $108$-cell stress grid,
$H\in\{0.07,0.1,0.2,0.3\}$, $\eta\in\{1,1.5,1.9\}$, $\rho\in\{-0.4,-0.7,-0.9\}$,
$T\in\{0.5,1,2\}$, against a $2^{18}$-path reference with $0.54$\,bp mean standard error,
it averages $0.51$\,bp (maximum $2.6$\,bp, at $H=0.07$, $\rho=-0.9$) and is within two
reference standard errors in $90$ cells; the error grows towards small $H$ and strong
correlation, $1.07$\,bp on average at $H=0.07$ against $0.17$\,bp at $H=0.3$.

\paragraph{The skeleton ladder on the Volterra field.}
At $\rho=0$ the characteristic function,
\[
  \varphi(u)=\EE\Bigl[\exp\Bigl(-w\,v_0\Delta t\sum_ke^{Y_k}\Bigr)\Bigr],
\]
has exactly the form of \eqref{eq:gamma-melon} with the Volterra covariance in place of the
OU one, so the evaluated hierarchy of \Cref{sec:diagrams} applies directly to the rough
field. \Cref{tab:rb-ladder} scores it against a conditional Black--Scholes RQMC average
on the same grid, whose standard error is below $0.25$\,bp. The 2PI Hartree level is
already within $1$--$15$\,bp, where the two-cumulant closure is $75$--$240$\,bp off; the
melon level reaches $0.12$\,bp in the mild and $1.2$\,bp in the moderate regime,
deterministically and with no sampling of the field. In the extreme regime it improves
only from $14.5$ to $9.5$\,bp, the plateau of \Cref{sec:freezing}. At $\rho\neq0$ the
leverage integral $\int e^{Y/2}\dd W^v$ cannot be localised as in \eqref{eq:ito-skew},
because the field is not Markov; the skeletons then couple the field to the driver
increments at different sites, and we have not evaluated them. At nonzero correlation
the composite closure takes their place: it conditions on the field's leading modes
instead of expanding around a Gaussian, and reaches the reference noise in all three
regimes, including the extreme one where the melon level plateaus.

\begin{table}[t]
\centering
\begin{tabular}{@{}lrrr@{}}
\toprule
\textbf{Method ($\rho=0$)} & \textbf{Mild} & \textbf{Moderate} & \textbf{Extreme} \\
\midrule
Two-cumulant Gaussian characteristic function & 75 & 129 & 240 \\
2PI Hartree (Gaussian variational) & 1.11 & 4.82 & 14.5 \\
2PI melon level & 0.12 & 1.22 & 9.45 \\
\midrule
Reference two standard errors & 0.01 & 0.13 & 0.23 \\
\bottomrule
\end{tabular}
\caption{Rough Bergomi at $\rho=0$: implied-vol RMSE (bp, $K/F\in[0.85,1.15]$) of the
evaluated skeleton levels against a conditional Black--Scholes RQMC average on the same
grid ($100$ steps per year at midpoints, $2^{17}$ Sobol points $\times$ $8$ scrambles,
PCA order). Regimes as in \Cref{tab:rbergomi} with $\rho$ set to zero.}
\label{tab:rb-ladder}
\end{table}

\begin{table}[t]
\centering
\begin{tabular}{@{}lrrr@{}}
\toprule
\textbf{Method} & \textbf{Mild} & \textbf{Moderate} & \textbf{Extreme} \\
\midrule
Two-cumulant Gaussian characteristic function & 186 & 405 & 733 \\
Four-cumulant NIG approximation & 5 & 29 & 56 \\
Conditional-field Gauss--Hermite & 3 & 23 & 78 \\
PCA-resolved conditional RQMC, 16k nodes & $0.9\pm0.1$ & $2.1\pm0.4$ & $2.7\pm0.9$ \\
PCA-resolved conditional RQMC, 32k nodes & $0.9\pm0.0$ & $1.5\pm0.4$ & $2.2\pm0.5$ \\
\midrule
\multicolumn{4}{@{}l}{\emph{against the $2^{24}$-path conditional reference, $N=200$}} \\
PCA-resolved conditional RQMC, 16k nodes & $0.18\pm0.07$ & $2.00\pm0.44$ & $2.85\pm0.99$ \\
PCA-resolved conditional RQMC, 32k nodes & $0.11\pm0.03$ & $1.35\pm0.38$ & $2.42\pm0.59$ \\
Composite closure, two modes (deterministic) & $0.002$ & $0.058$ & $0.115$ \\
\bottomrule
\end{tabular}
\caption{Rough Bergomi implied-volatility RMSE (bp) on the fixed grid
$K/F\in[0.85,1.15]$. Upper block: versus Sobol Monte Carlo ($2^{19}$ paths, same time
grid). Lower block: versus a conditional Monte Carlo on the same $200$-step grid, which
draws $(Y,\Delta W^v)$ exactly, prices each path by Black's formula and uses
control variates and antithetic pairs, for a standard error of at most $0.12$\,bp with
$2^{24}$ paths; seven strikes. The first two rows
are cumulant closures of the log-return; the conditional-field Gauss--Hermite and
PCA-resolved RQMC rows condition on the full discrete field. The RQMC rows report the mean $\pm$ cross-scramble standard error over
eight independent Sobol scrambles (a displayed $\pm0.0$ denotes $<0.05$\,bp). These $\pm$ values are
cross-scramble standard errors of the conditional-RQMC estimator and do not include the uncertainty
of the finite-path Monte-Carlo reference; the upper-block RMSE figures should therefore be read as
agreement to the reference noise floor rather than as demonstrated sub-basis-point accuracy,
which the lower block, with a much less noisy reference, resolves. The composite-closure
RMSEs are comparable to that reference's standard error ($0.013$, $0.04$ and $0.10$\,bp in
the three regimes) and so measure agreement approximately at its resolution. Regimes:
Mild ($H{=}0.20$, $\eta{=}1.0$, $\rho{=}-0.40$, $v_0{=}0.04$, $T{=}1$\,y), Moderate
($H{=}0.10$, $\eta{=}1.5$, $\rho{=}-0.70$, $v_0{=}0.04$, $T{=}1$\,y), Extreme
($H{=}0.07$, $\eta{=}1.9$, $\rho{=}-0.90$, $v_0{=}0.04$, $T{=}2$\,y).}
\label{tab:rbergomi}
\end{table}

\begin{figure}[t]
\centering
\includegraphics[width=\textwidth]{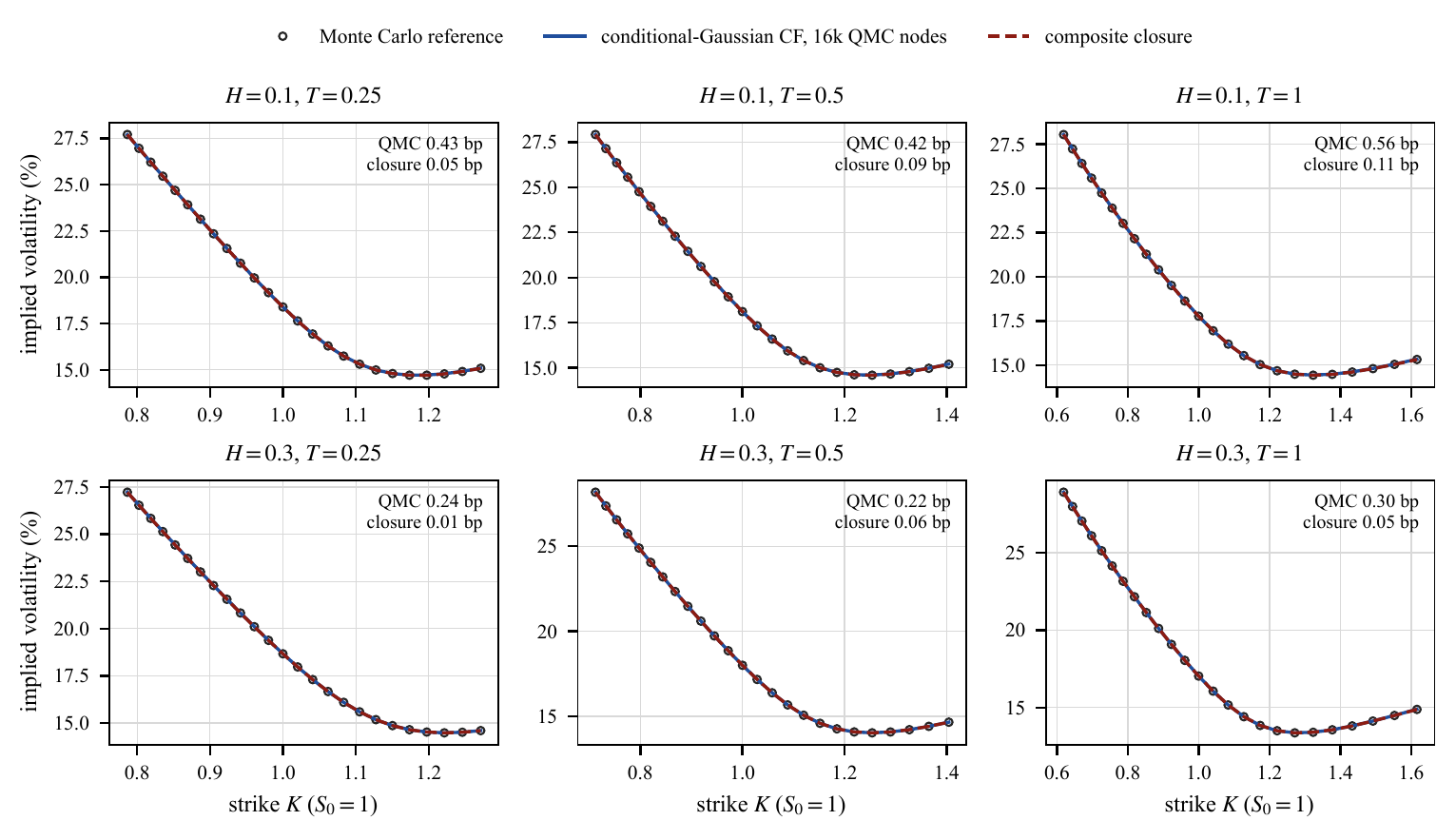}
\caption{Rough Bergomi smiles ($v_0=0.04$, $\rho=-0.7$, $\eta=1.5$, $N=200$) across
$H\in\{0.1,0.3\}$ (rows) and $T\in\{0.25,0.5,1\}$\,y (columns). Markers: conditional
Monte-Carlo reference, $2^{24}$ paths (standard error at most $0.11$\,bp). Solid:
PCA-resolved conditional RQMC, $16$k nodes, one scramble ($0.22$--$0.56$\,bp panel RMSE).
Dashed: composite closure ($0.01$--$0.11$\,bp, worst strike $0.31$\,bp). Strikes span
$\pm2.4$ at-the-money standard deviations, $K/F\in[0.62,1.62]$ at $T=1$.}
\label{fig:rbergomi-grid}
\end{figure}

\begin{figure}[t]
\centering
\includegraphics[width=0.86\textwidth]{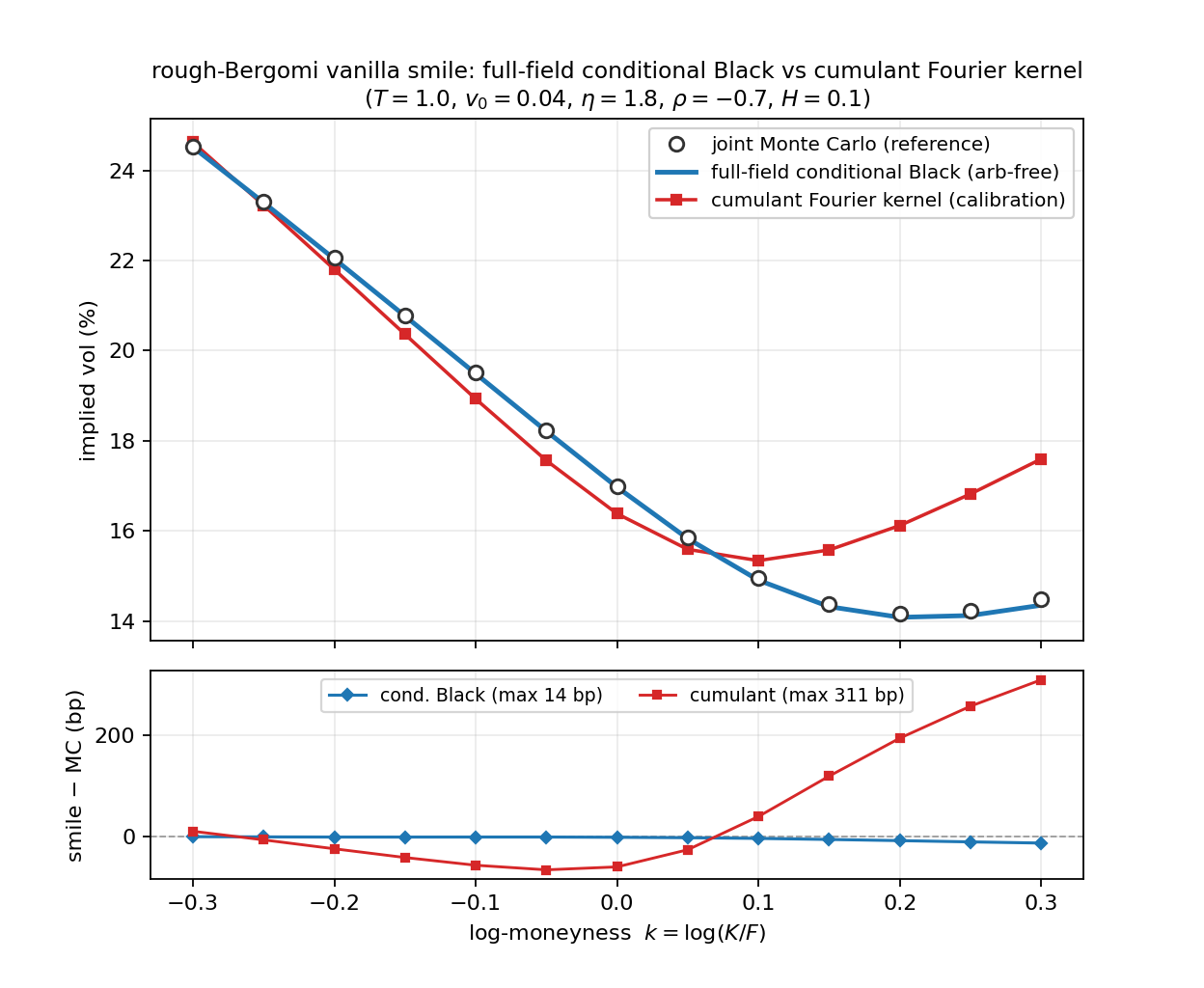}
\caption{Rough-Bergomi vanilla smile ($T=1$, $v_0=0.04$, $\eta=1.8$, $\rho=-0.7$,
$H=0.1$). The full-field conditional Black average (blue) is manifestly
arbitrage-free and matches Monte Carlo to $\sim$14\,bp; the truncated-cumulant
kernel (red), the fast calibration route, develops several-hundred-bp wing errors
where its density loses positivity.}
\label{fig:rbergomi-vanilla}
\end{figure}

The wing residual quoted in vol points (tens of bp in the deepest wing) is largely a
vega artefact: converted back to price it is flat and small (RMSE $0.18$\,bp of spot
at one year in the strong-coupling cell), inside the Monte-Carlo band, because vega
collapses by two orders of magnitude into the wing. The benchmark shares the time grid
between the pricer and the reference on purpose: it measures the pricer's error for
the discretised model, and the common $O(N^{-(H+1/2)})$ discretisation error against
the continuum model is reported separately (\supp{S7}; at $H=0.1$ it is the larger of
the two in the deep call wing). The vol-point figures quoted here and in the abstract
are therefore agreement with the discretised model, not accuracy against the
continuum limit. A free first-cumulant control
variate (the Level-1 mean, already computed) cuts the outer standard error by
$2.2$--$2.7\times$ where the error lives; the diagnostics are collected in \supp{S7}.
An arbitrage-free cross-check is available at no modelling cost: the vanilla is the
$t_1=0$ forward-start smile, so $C_0(k)=\EE_{\text{field}}[\mathrm{Black}(k)]$ is a
convex average of genuine Black prices and manifestly arbitrage-free. It tracks the
reference to $\sim$14\,bp across the strike range (a few bp near the money, growing
into the deep wings), whereas the truncated-cumulant kernel develops
several-hundred-bp wing errors (\Cref{fig:rbergomi-vanilla}). That kernel is fast,
but the exponential of a cumulant polynomial truncated beyond second order is never
the characteristic function of a probability law~\cite{Marcinkiewicz1939}: it fails
Bochner's positive-definiteness condition, so its Fourier inversion returns a
``density'' that turns negative in the wings, which is exactly where the errors
appear. The composite closure is arbitrage-free for the same reason as the field
average: it is a mixture of Black prices with non-negative Gauss--Hermite weights.

\subsection{Rough SABR}
\label{sec:rsabr}
\begin{figure}[t]
\centering
\includegraphics[width=\textwidth]{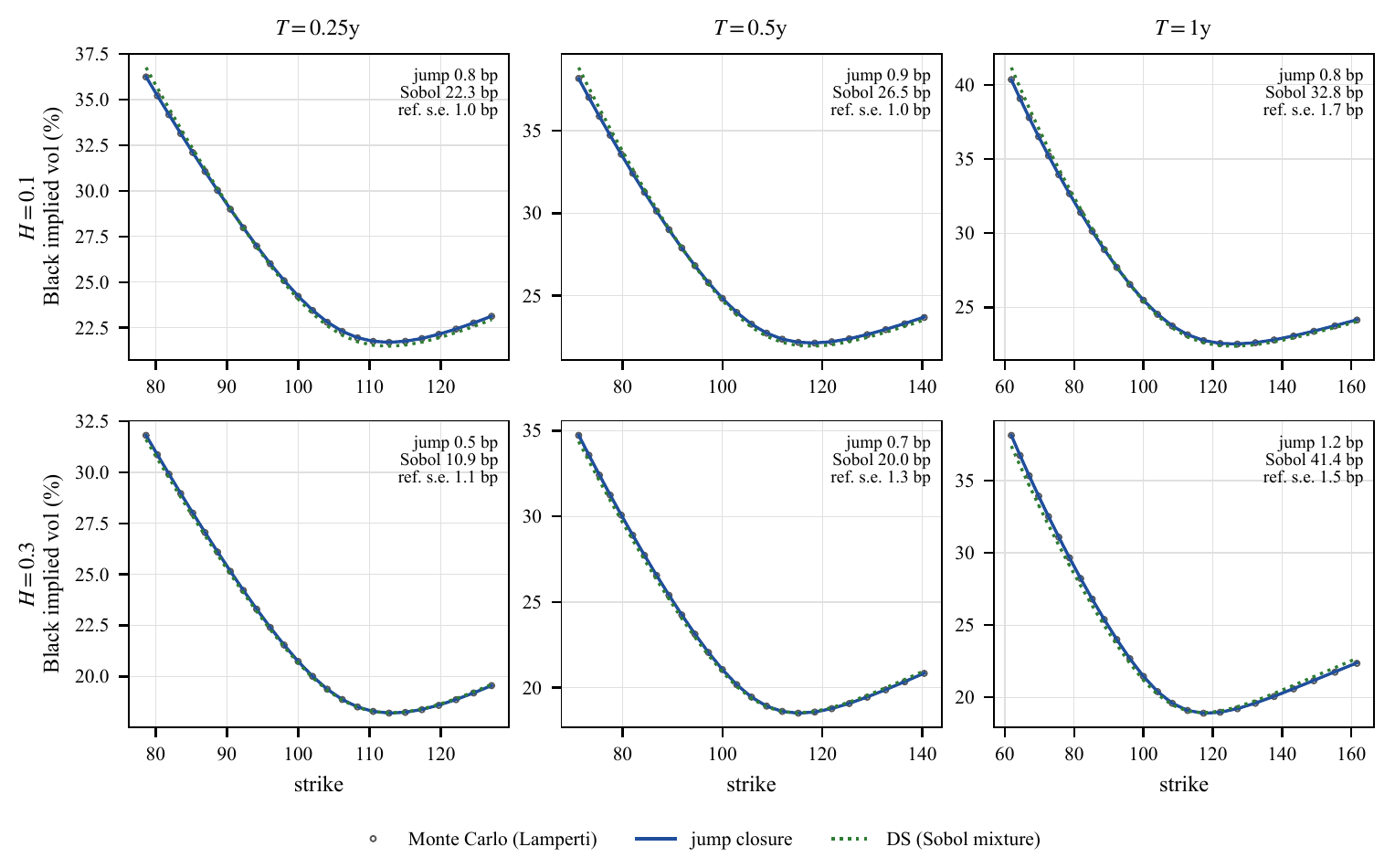}
\caption{Rough SABR smiles ($F_0=100$, ATM $\approx20\%$, $\beta=0.5$, $\rho=-0.5$,
$\eta=1$) across $H\in\{0.1,0.3\}$ (rows) and $T\in\{0.25,0.5,1\}$\,y (columns), strikes
spanning $\pm2.4$ at-the-money standard deviations. Markers and band: Lamperti-space
Monte-Carlo reference with $\pm2$ standard errors. Solid: jump closure ($0.5$--$1.2$\,bp
panel RMSE). Dotted: conditional-field RQMC ($11$--$41$\,bp). Lacking a
finite-maturity Hagan analogue~\cite{FukasawaGatheral2022}, both are shown directly
against the reference.}
\label{fig:rsabr-grid}
\end{figure}

Rough SABR combines the Volterra kernel with the CEV map---the rough counterpart of
\Cref{sec:sabr}. Conditional on the volatility path the forward is a pure CEV
diffusion characterised by the integrated variance $V=\int_0^T\sigma_t^2\,\dd t$ and
the correlation-induced Lamperti drift; the two-time conditional-field RQMC samples
the field by PCA-ordered Sobol, computes the effective start and
$\sigma_{\text{eff}}=\sqrt{(1-\rho^2)V/T}$ by Gaussian regression, and prices from
the exact CEV Green's function with a martingale renormalisation: a rescaling of the
effective starts for $H\ge0.2$, and for $H<0.2$ a multiplication of the prices by
$\lambda=F_0/\EE[F^{\mathrm{eff}}]$, within $0.7\%$ of one in the reported cells, which
keeps them convex and decreasing in the strike but leaves the implied measure with mass
$\lambda$ rather than one. As in ordinary
SABR the conditional-CEV reduction is an approximation---the $\rho$-displacement is
the direct analogue of the Islah displacement and the booster ports because its
inputs are Gaussian-bridge functionals---and the displacement-timing error is the
main reason the residual reaches tens of basis points in the strong-coupling wing.
The jump closure of \Cref{sec:sabr} removes it unchanged: SABR is the case $H=\tfrac12$
of the same construction, and only the mode functionals and the conditional jump means
are computed on the Volterra field. Rough SABR also needs a sharper reference than the
first version of this article used. Its Euler-in-$F$ scheme with $400$ steps moved by
$3$--$12$\,bp between $400$ and $800$ steps in four of the six cells of
\Cref{fig:rsabr-grid}, so we now simulate the forward in the Lamperti variable, with the
same principal-component Volterra field and Brownian-bridge perpendicular noise, $2^{15}$
paths $\times$ $128$ scrambles ($256$ in the six figure cells) and $400$ steps, for a mean
standard error of $1.1$\,bp (maximum $5.0$\,bp).

\begin{table}[t]
\centering
\small
\begin{tabular}{@{}lrrrrrrrr@{}}
\toprule
 & \multicolumn{4}{c}{\textbf{conditional-field RQMC}} & \multicolumn{4}{c}{\textbf{jump closure}} \\
\cmidrule(lr){2-5}\cmidrule(lr){6-9}
 & \multicolumn{2}{c}{$H=0.1$} & \multicolumn{2}{c}{$H=0.3$} & \multicolumn{2}{c}{$H=0.1$} & \multicolumn{2}{c}{$H=0.3$} \\
$\rho$ & mean & max & mean & max & mean & max & mean & max \\
\midrule
$-0.5$ & 30.8 & 56.9 & 16.8 & 34.0 & 0.86 & 1.37 & 0.47 & 0.66 \\
$0$    &  1.5 &  2.9 &  0.4 &  0.8 & 1.08 & 1.70 & 0.75 & 1.24 \\
$+0.25$& 17.0 & 28.1 &  1.2 &  5.3 & 1.49 & 1.96 & 0.88 & 1.70 \\
\bottomrule
\end{tabular}
\caption{Rough SABR implied-vol RMSE (bp, 21 strikes $75$--$125$) against the
Lamperti-space Monte-Carlo reference, mean and maximum over the $12$ cells
$\beta\in\{0.3,0.5,0.7\}\times\eta\in\{0.5,1\}\times T\in\{0.5,1\}$ of each
$(\rho,H)$ pair. The RQMC column uses a $200$-point field and $4{,}096$ Sobol points; the
jump closure uses seven Lobatto jumps.}
\label{tab:rsabr}
\end{table}

Against this reference the conditional-field RQMC averages $11.3$\,bp, and up to
$57$\,bp at $\rho=-0.5$, $H=0.1$; raising its Sobol count from $4{,}096$ to $32{,}768$
barely changes this, so the error is the single displaced start rather than the field
quadrature. The jump closure stays below $2$\,bp in all $72$ cells of \Cref{tab:rsabr},
with mean $0.92$\,bp and worst single strike $4.2$\,bp. On the wider strike strips of
\Cref{fig:rsabr-grid}, two cells at $T=0.25$ and four at grid parameters, it is within
$1.2$\,bp against $11$--$41$\,bp for the RQMC average; at $\rho=0$, where the RQMC average is exact
up to the field discretisation, the two agree to within the reference noise.

\paragraph{An exactly-solvable member: quadratic Gaussian Volterra variance.}
One case escapes closure entirely. If the spot variance is a \emph{quadratic}
function of a \emph{free} Gaussian Volterra field, $v_t=a(Z_t-b)^2+c$, the integrated
variance is a quadratic form in a Gaussian vector and its moment-generating function
is a single determinant, $\EE[e^{\theta Y}]=\exp(\theta cT-\tfrac12\log\det
M_\theta+\theta a\Delta t\,b^2\mathbf 1^\top M_\theta^{-1}\mathbf 1)$ with
$M_\theta=I-2\theta a\Delta t\,C$---no closure at all. This frozen-feedback analogue
of quadratic rough Heston~\cite{GatheralJusselinRosenbaum2020} is the lattice image
of the Fredholm-determinant transforms of Abi~Jaber~\cite{AbiJaberWishart2022,
AbiJaberGaussian2022}. Its admissible strip is exact from a single eigenvalue,
$\theta_{\max}=1/(2a\Delta t\lambda_{\max}(C))$, and is wide and strictly
positive---in contrast to the empty positive strip of the exponential Bergomi
functional---and the determinant matches a $4\times10^6$-path reference to its own
noise floor at $\sim$$90\times$ lower cost. It is a useful exactly-solvable
forward-variance marginal and control variate; details are in \supp{S8}.

\subsection{Multi-asset consistency: the FX triangle}
\label{sec:fxtriangle}
\begin{figure}[t]
\centering
\includegraphics[width=\textwidth]{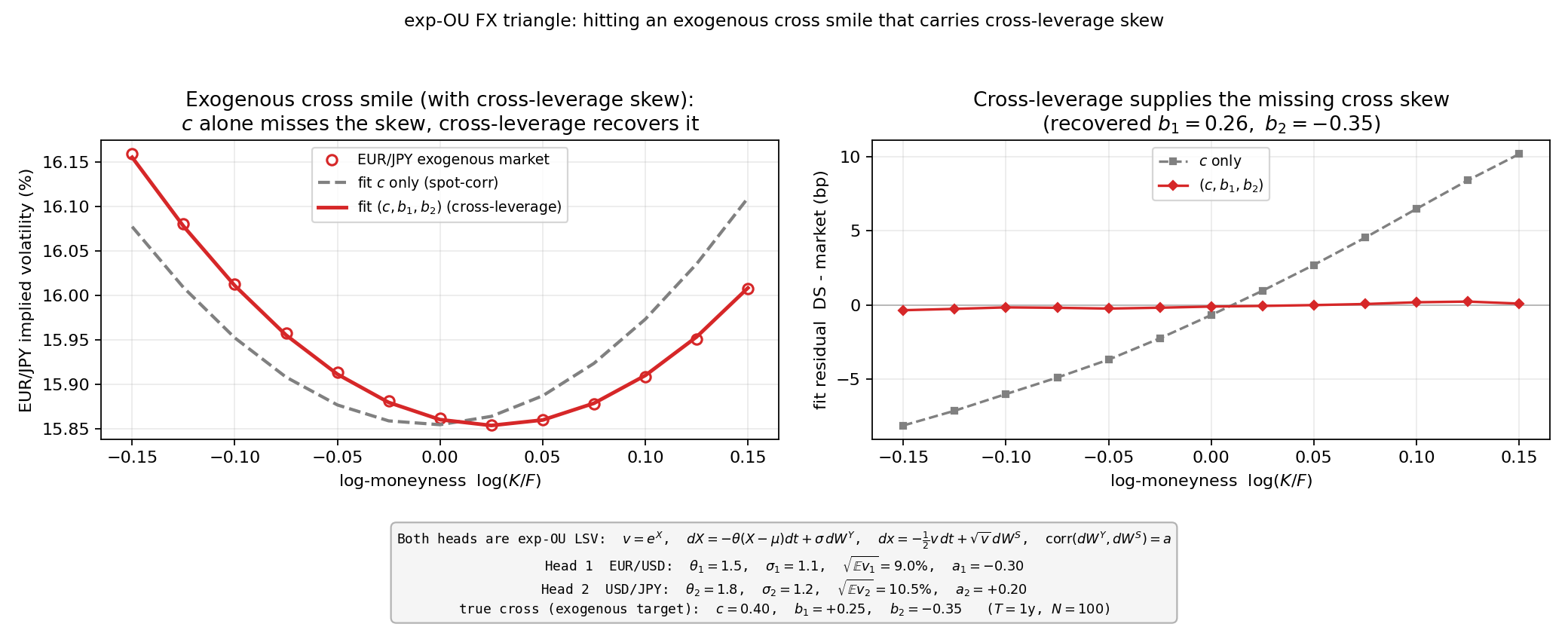}
\caption{exp-OU FX triangle against a synthetically generated exogenous EUR/JPY target
smile carrying cross-leverage skew. \emph{Left:} the exogenous cross with the
spot-correlation-only fit ($c$ alone)
and the cross-leverage fit $(c,b_1,b_2)$. \emph{Right:} the residual---$c$ alone
rotates from $-8$ to $+10$\,bp, cross-leverage flattens it to $\pm0.5$\,bp. The heads
are reproduced to $0.3$--$0.5$\,bp and are invariant to the cross knobs.}
\label{fig:fxtriangle}
\end{figure}

Multiple assets impose a constraint no single-name smile can see: three FX pairs form
a no-arbitrage triangle, $z=x+y$ in log-returns, so the cross marginal is determined
by the joint law of the heads. The conditional-Gaussian representation enforces this
exactly, because the linear relation is preserved pathwise: conditional on the
volatility fields the log-returns are jointly Gaussian, so the cross marginal is a
closed field average of a Black value, arbitrage-free whenever the joint correlation
structure is admissible and \emph{constrained} rather than freely fitted. The heads
depend only on their own leverage; the couplings---spot--spot correlation $c$,
co-roughness $g$ (rough only), and cross-leverage $(b_1,b_2)$---move the cross without
disturbing them, with $(b_1,b_2)$ the dedicated cross-skew handle. Fed a
synthetically generated exogenous EUR/JPY target smile carrying independent
cross-leverage, spot correlation alone matches the
level but leaves a skew residual ($5.8$\,bp exp-OU, $9.3$\,bp rough), which the full
$(c,g,b_1,b_2)$ fit collapses to $0.2$/$1.5$\,bp while recovering the true couplings
(\Cref{tab:fxtri}, \Cref{fig:fxtriangle}). The construction crosses the
Markovian/rough fork with only the field sampler changing.

\begin{table}[t]
\centering
\begin{tabular}{@{}lrr@{}}
\toprule
\textbf{Cross-smile RMSE vs.\ exogenous target (bp)} & \textbf{exp-OU} & \textbf{rough Bergomi} \\
\midrule
Head marginals (EUR/USD, USD/JPY) & 0.3,\ 0.5 & 1.5,\ 3.2 \\
Cross, level only$^{a}$ & 5.8 & 9.3 \\
Cross, with cross-leverage $(b_1,b_2)$ & 0.2 & 1.5 \\
\bottomrule
\end{tabular}
\caption{FX-triangle fit to a synthetically generated exogenous EUR/JPY target smile
carrying independent cross-leverage. $^{a}$Level-only means $c$ (exp-OU) or $(c,g)$
(rough). The heads are
reproduced by construction; the cross-leverage handle closes the residual.}
\label{tab:fxtri}
\end{table}

\section{The Evaluated Skeleton Hierarchy}
\label{sec:diagrams}
The Luttinger--Ward functional $\Phi$ is the sum of two-particle-irreducible
skeletons built from the dressed propagator; its first terms are the Hartree bubble,
the sunset and the basketball, and the self-energy $\Sigma=-2\,\delta\Phi/\delta G$ opens
one propagator line in each. For exponential vertices these skeletons can be summed in
closed form, which turns the hierarchy from a classification into a computation.

\begin{center}
\begin{tikzpicture}[baseline=(current bounding box.center), scale=0.9]
  \begin{scope}[shift={(0,0)}]
    \node[vtx] (v1) at (0,0) {};
    \draw[xprop dressed] (v1) to[out=60,in=120,looseness=5] (v1);
    \node[above=20pt] at (v1) {\footnotesize\textbf{Hartree}};
    \node[above=10pt] at (v1) {\footnotesize (1-loop)};
    \node[below=6pt] at (v1) {\footnotesize $\tfrac12 V_0\,G(0)$};
  \end{scope}
  \begin{scope}[shift={(5,0)}]
    \node[vtx3] (a) at (-0.8,0) {};
    \node[vtx3] (b) at (0.8,0) {};
    \draw[xprop dressed] (a) to[out=50,in=130] (b);
    \draw[xprop dressed] (a) -- (b);
    \draw[xprop dressed] (a) to[out=-50,in=-130] (b);
    \node[above=16pt] at (0,0.45) {\footnotesize\textbf{Sunset}};
    \node[above=6pt] at (0,0.45) {\footnotesize (2-loop)};
    \node[below=8pt] at (0,-0.5) {\footnotesize $\propto V_0^2[G]^3$};
  \end{scope}
  \begin{scope}[shift={(10.5,0)}]
    \node[vtx4] (c) at (-0.8,0) {};
    \node[vtx4] (d) at (0.8,0) {};
    \draw[xprop dressed] (c) to[out=60,in=120] (d);
    \draw[xprop dressed] (c) to[out=25,in=155] (d);
    \draw[xprop dressed] (c) to[out=-25,in=-155] (d);
    \draw[xprop dressed] (c) to[out=-60,in=-120] (d);
    \node[above=16pt] at (0,0.55) {\footnotesize\textbf{Basketball}};
    \node[above=6pt] at (0,0.55) {\footnotesize (3-loop)};
    \node[below=8pt] at (0,-0.7) {\footnotesize $\propto V_0^2[G]^4$};
  \end{scope}
\end{tikzpicture}
\end{center}

\paragraph{The functional.}
Discretise the log-volatility fluctuation on the pricing grid, $\phi\sim\mathcal N(0,K)$
on sites $k=1,\dots,n$, and write the characteristic function as
$\varphi(u)=\EE[e^{-\sum_k s_k(\phi_k)}]$ with local vertices $s_k$. At $\rho=0$ in
exp-OU and rough Bergomi, $s_k(\phi)=w\,\Delta t\,e^{\mu_k+\phi}$ with
$w=\tfrac12(\ii u+u^2)$. For a Gaussian of mean $m$ and covariance $G$, let
$D_k(n)=\EE[\partial^n s_k]$ at $\mathcal N(m_k,G_{kk})$; for $s_k=c_ke^{a\phi}$ this is
$D_k(n)=a^nc_ke^{am_k+a^2G_{kk}/2}$. The 2PI functional truncated at two vertices is
\begin{equation}\label{eq:gamma-melon}
  \begin{aligned}
  \Gamma[m,G]&=\tfrac12m^{\!\top}K^{-1}m+\tfrac12\operatorname{Tr}(K^{-1}G)
  -\tfrac12\log\det(K^{-1}G)-\tfrac n2\\
  &\quad+\sum_kD_k(0)-\tfrac12\sum_{n\ge3}\sum_{k,l}\frac{G_{kl}^n}{n!}D_k(n)D_l(n),
  \end{aligned}
\end{equation}
and $\log\varphi(u)=-\Gamma$ at its stationary point. The first four terms are the
Gaussian part; $D_k(0)$ is the Hartree vertex, which carries every self-loop through the
exact moment-generating function \eqref{eq:resum}; the last sum is the exchange of $n$
dressed lines between two vertices. The terms $n=1$ and $n=2$ are omitted because they
are one- and two-particle reducible: the mean equation and the $\operatorname{Tr}\log$
rings already generate them. For exponential vertices the remaining sum is the melon
kernel,
\begin{equation}\label{eq:melon}
  \sum_{n\ge3}\frac{(a_ka_lG_{kl})^n}{n!}=e^{a_ka_lG_{kl}}-1-a_ka_lG_{kl}-\tfrac12(a_ka_lG_{kl})^2,
\end{equation}
containing the sunset ($n=3$), the basketball ($n=4$) and every higher melon. Setting it
to zero leaves the Hartree level, which is the Gaussian variational approximation of
the path measure. Stationarity, with $\partial_{m_k}D_k(n)=D_k(n+1)$ and
$\partial_{G_{kk}}D_k(n)=\tfrac12D_k(n+2)$ (Price's theorem), gives the gap equations
\begin{equation}\label{eq:gap-melon}
  m=-K\bigl[D(1)-M^{(1)}\bigr],\qquad
  G^{-1}=K^{-1}+\operatorname{diag}\bigl(D(2)-M^{(2)}\bigr)-P,
\end{equation}
with $M^{(j)}_k=\sum_{n\ge3}D_k(n+j)\sum_l\frac{G_{kl}^n}{n!}D_l(n)$ and
$P_{kl}=\sum_{n\ge3}\frac{G_{kl}^{n-1}}{(n-1)!}D_k(n)D_l(n)$, solved by damped fixed-point
iteration with continuation in $u$. Every level is $\Phi$-derivable in the sense
of~\cite{BaymKadanoff1961}: the mean and covariance are stationary points of one
functional, so a first-order error in them enters $\log\varphi$ only at second order.

\paragraph{The fork is a property of the level.}
The melon self-energy $P_{kl}$ couples every pair of times. In exp-OU $K^{-1}$ is
tridiagonal and the Hartree self-energy is diagonal, which is why the Level-1 gap
equation reduces to equal-time ODEs; at the melon level the dressed inverse propagator
is dense even for this Markovian model. The next skeleton therefore needs the two-time
machinery of the rough branch, so the Markovian/rough fork is decided by the model and
the truncation level together.

\paragraph{Correlation.}
At $\rho\neq0$ the exp-OU exponent contains $\ii u\rho\int e^{X/2}\,\dd W^X$, which is not
a function of the field at single times. It\^o's formula for $g(X)=(2/\sigma)e^{X/2}$
removes the stochastic integral,
\begin{equation}\label{eq:ito-skew}
  \int_0^Te^{X_t/2}\,\dd W^X_t=\tfrac2\sigma\bigl(e^{X_T/2}-e^{X_0/2}\bigr)
  +\int_0^Te^{X_t/2}\Bigl(\tfrac\theta\sigma(X_t-\mu)-\tfrac\sigma4\Bigr)\dd t,
\end{equation}
so every vertex is again local, of the form $c\,\phi^pe^{a\phi}$ with $p\in\{0,1\}$ and
$a\in\{\tfrac12,1\}$, plus two endpoint vertices; the spot's orthogonal leg contributes
$-\tfrac12u^2(1-\rho^2)\int v\,\dd t$. The derivatives $D_k(n)$ keep closed forms, for
example $\EE[\partial^n(\phi e^{a\phi})]=e^{am+a^2v/2}\bigl(a^n(m+av)+na^{n-1}\bigr)$, and the
same functional and gap equations apply.

\paragraph{Three vertices.}
The next skeletons join three vertices by bundles of $n_{12},n_{23},n_{31}\ge0$ lines,
with the two-particle-irreducibility condition that no vertex can be separated by
cutting two lines, $n_{12}+n_{31}\ge3$ and cyclically. Grouping the line counts into the
classes $0,1,2,\ge3$ turns each admissible pattern into a matrix product,
$\operatorname{tr}(D\,E_{c_1}D\,E_{c_2}D\,E_{c_3})$ with $D=\operatorname{diag}(D(0))$ and
$E_c\in\{1,G,\tfrac12G^2,e^G-1-G-\tfrac12G^2\}$ elementwise, contributing
$+\tfrac16\kappa_3$ to $\Gamma$. By stationarity it can be evaluated at the melon solution
without re-solving.

\begin{table}[t]
\centering
\small
\begin{tabular}{@{}lrrrr@{}}
\toprule
\textbf{Cell} & \textbf{2PI Hartree} & \textbf{melon} & \textbf{+ three-vertex} & \textbf{SCG} \\
\midrule
$\sigma=1.5$, $\rho=0$, $T=0.25$--$2$ & $0.36$--$0.81$ & $0.01$--$0.04$ & $\le0.01$ & $0.05$--$0.19$ \\
$\sigma=1.5$, $\rho=-0.7$, $T=0.25$--$2$ & $3.1$--$4.8$ & $0.09$--$0.11$ & --- & --- \\
$\sigma=1.5$, $\rho=-0.4$, $T=0.25$--$2$ & $0.9$--$1.4$ & $0.02$--$0.04$ & --- & --- \\
$\sigma=1.5$, $\rho=+0.4$, $T=0.25$--$2$ & $0.9$--$2.7$ & $0.02$--$0.03$ & --- & --- \\
$\sigma=3$, $\rho=0$, $T=0.5$--$5$ & $14.6$--$24.8$ & $3.0$--$9.5$ & $1.7$--$14.5$ & $5.0$--$10.3$ \\
$\theta=1$, $\rho=0$, $T=1$--$10$ & $7.8$--$14.2$ & $2.3$--$3.7$ & $0.4$--$3.4$ & $0.7$--$4.1$ \\
$\sigma=3$, $\rho=-0.7$, $T=1,5$ & $39.8,\ 161.9$ & $3.6,\ 40.7$ & --- & --- \\
\bottomrule
\end{tabular}
\caption{Exp-OU implied-vol RMSE (bp, strikes $80$--$120$) of the evaluated skeleton
levels against the exact tower recursion on the same time grid ($\theta=3$, $\mu=-2$
unless stated). At $\rho\neq0$ the tower includes the endpoint weights of
\eqref{eq:ito-skew}; because that discrete functional has a large time-step error of its
own, the continuous-time SCG is not scored there. Dashes mark levels not evaluated.}
\label{tab:ladder}
\end{table}

\paragraph{What the ladder shows.}
\Cref{tab:ladder} collects the tests. The melon level improves on Hartree in every cell,
by $20$--$130\times$ at moderate coupling and $3$--$11\times$ at strong coupling, and its
gain is as large at $\rho\neq0$ as at $\rho=0$. At strong coupling the levels alternate in
sign (Hartree high, melon low, three-vertex high) and the three-vertex correction helps
at long maturities but hurts at short maturity, where the pricing transform needs large
$|u|$: the skeleton series behaves as an asymptotic series, as it must when the
integrated variance has no positive exponential moment (\Cref{sec:freezing}). Two
qualifications apply. The 2PI Hartree level and the SCG of \Cref{sec:expou} are
different Gaussian closures---the first projects the path measure, the second the
Feynman--Kac wave function under the tilted measure---and the SCG is the more accurate
of the two at Level~1; the skeleton corrections are built on the former. And the melon
level costs a dense $n\times n$ solve per frequency, $4$--$150$\,s per smile in the
single-core Python reference against milliseconds for the SCG, so it serves as a check on
the hierarchy and as a deterministic reference rather than as a calibration engine. On
the rough-Bergomi Volterra field the same functional applies unchanged
(\Cref{sec:rbergomi}).

The quadratic Gaussian Volterra model marks where the tower terminates for a
structural reason: with the variance a quadratic form in a Gaussian field the
exponent stays quadratic and the average is one determinant, no skeleton hierarchy to
truncate---and, with no closure error to absorb it, the field-covariance
discretisation becomes the binding constraint, so making one layer exact simply
migrates the bottleneck down the hierarchy.


\section{Fast Exotics via the Two-Time Propagator}
\label{sec:exotics}
Off the diagonal, the same conditional-Gaussian law prices path-dependent and
forward-looking payoffs directly. Conditional on the latent Gaussian volatility
field, the log-spot is a time-changed, drifted Gaussian process,
\begin{equation}\label{eq:condgauss}
  X_t = A_t + B_{c_t},\qquad
  A_t = -\tfrac12\!\int_0^t\! v_s\,\dd s + \rho\!\int_0^t\!\sqrt{v_s}\,\dd W^X_s,
  \qquad
  c_t = (1-\rho^2)\!\int_0^t\! v_s\,\dd s,
\end{equation}
with $B$ a Brownian motion independent of the field, so
$\Cov(X_t,X_s\mid\text{field})=c_{t\wedge s}$. Every payoff depending on $X$ at
finitely many dates, or on its running extremum, therefore has a closed conditional
law---Gaussian for a forward start, an image-propagator reflection expression for a
barrier (exact for a fixed boundary, numerical for the moving boundary that leverage
induces)---and its price is the field average $\EE_{\text{field}}[\,\cdot\,]$,
evaluated by the same PCA-ordered Sobol quadrature Part~I uses for the marginal.
Restricting \eqref{eq:condgauss} to the diagonal $c_T$ returns the vanilla; the
off-diagonal content prices the exotic, with neither a spot PDE nor spot-path
simulation. Only the sampler that draws $v$ changes between branches, so the
identical machinery carries from the exp-OU proxy to the native Gaussian--Volterra
field; we demonstrate the full exotic book on both.

\subsection{Markovian exotics: one-touch, reverse knock-out, forward-vol}
\label{sec:lsv-exotics}
\begin{figure}[p]
\centering
\includegraphics[width=\textwidth]{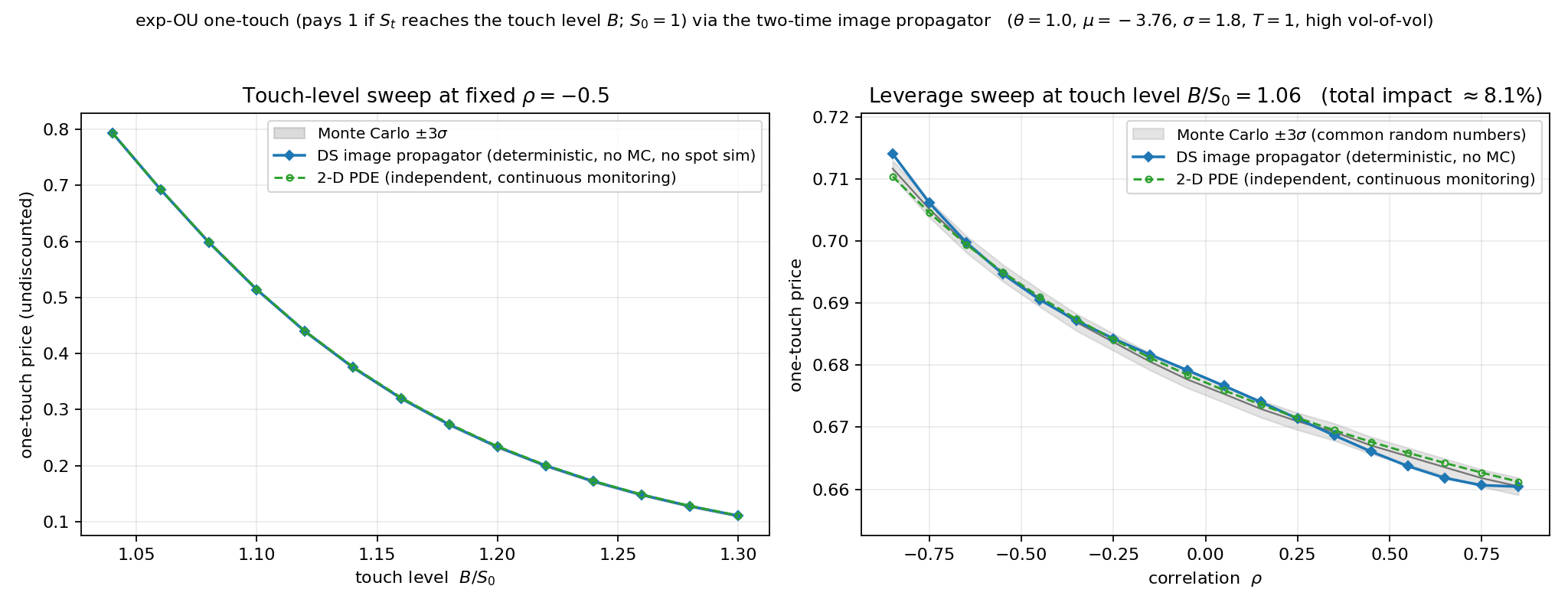}
\caption{exp-OU upside one-touch via the image propagator ($\theta=1$,
$\sigma=1.8$, base vol $\approx23\%$). \emph{Left:} price versus touch level at
$\rho=-0.5$. \emph{Right:} leverage sweep at touch level $B/S_0=1.06$, a monotone $\sim$8\%
correlation signal. The DS price (blue, no spot simulation) sits on the
Monte-Carlo $\pm3\sigma$ band; the green dashed curve is the independent
two-dimensional PDE, which exposes the image propagator's small discretisation tilt
on the far positive-$\rho$ wing.}
\label{fig:expou-onetouch}
\end{figure}

\begin{figure}[p]
\centering
\includegraphics[width=\textwidth]{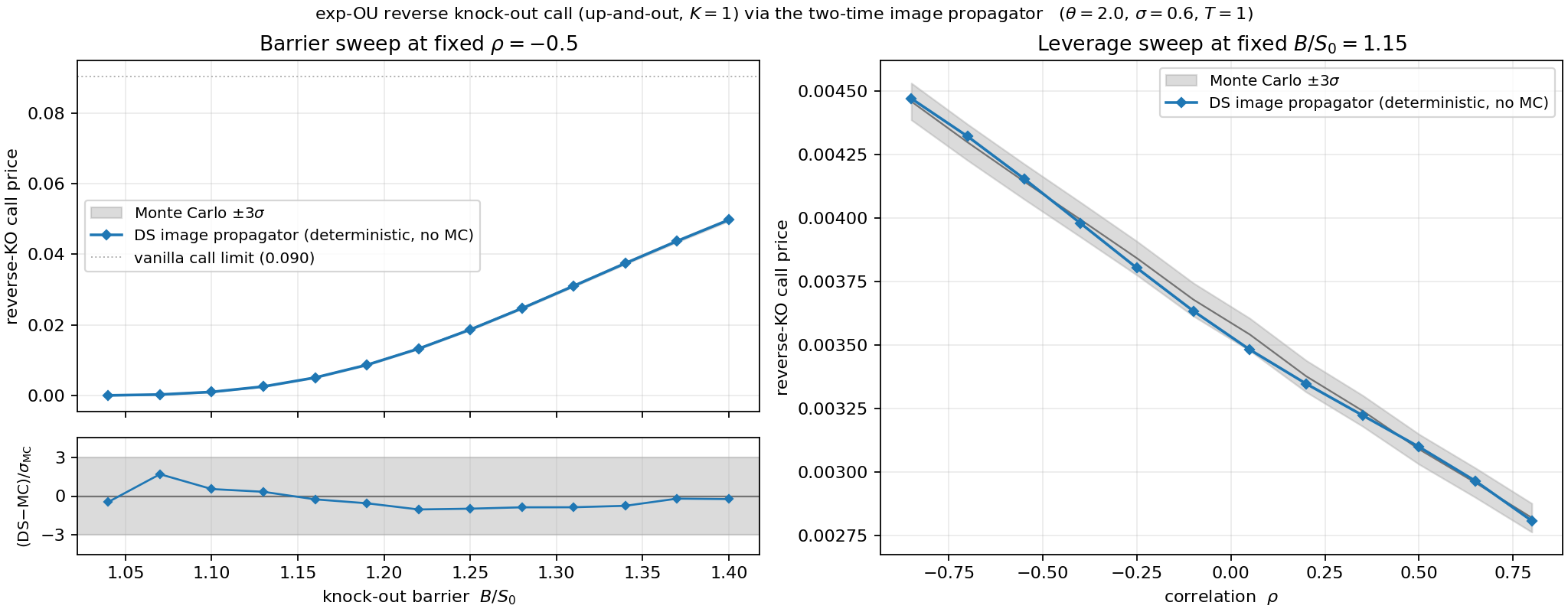}
\caption{exp-OU reverse knock-out call (up-and-out, $K=1$; $\theta=2$, $\sigma=0.6$).
\emph{Left:} price versus barrier at $\rho=-0.5$, rising towards the vanilla-call
limit (dotted); at this price scale the Monte-Carlo band is thinner than the line, so
the lower panel shows the residual DS$-$MC in units of the Monte-Carlo standard error,
inside $\pm2\sigma$ at every barrier. \emph{Right:} leverage sweep at $B/S_0=1.15$.
DS lies on the Monte-Carlo $\pm3\sigma$ band across both sweeps.}
\label{fig:expou-rko}
\end{figure}

\begin{figure}[p]
\centering
\includegraphics[width=0.82\textwidth]{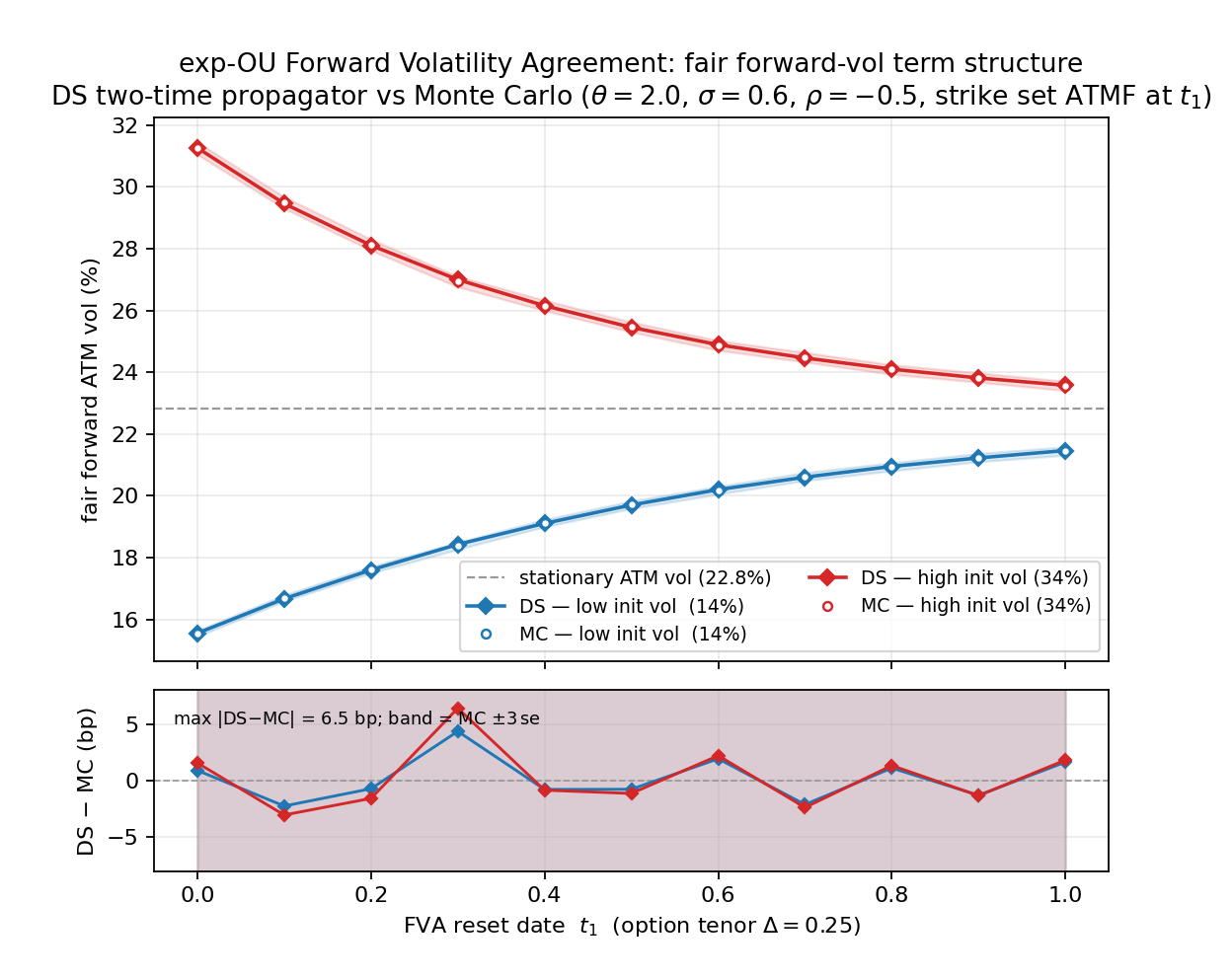}
\caption{exp-OU forward volatility agreement: fair forward at-the-money vol versus
reset date $t_1$ ($\Delta=0.25$; $\theta=2$, $\sigma=0.6$, $\rho=-0.5$). From a low
($14\%$) or high ($34\%$) initial vol the fair forward vol reverts toward the
stationary level (dashed); DS (filled diamonds) matches Monte Carlo (open circles)
throughout, the residual inside the $\pm3\sigma$ envelope with a maximum discrepancy
of $6.5$\,bp.}
\label{fig:expou-fva}
\end{figure}

The image propagator returns the terminal absorbed density on the survival set, so a
one-touch costs a touch probability and a barrier payoff the same propagation with a
different terminal accumulation. The upside one-touch carries a clean monotone
$\sim$8\% leverage signal in $\rho$, tracked to within a few Monte-Carlo standard
errors across $\rho\in[-0.85,0.85]$ and independently corroborated by a
two-dimensional Crank--Nicolson solve with a continuous absorbing wall---which shares
no machinery with the image propagator and localises the small residual to the
propagator's $O(\dd x)+O(1/M)$ discretisation (\Cref{fig:expou-onetouch}). The
reverse knock-out, worth most exactly where it is being extinguished, is the sharper
test: it carries a $\sim$46\% monotone leverage signal and stays on the Monte-Carlo
band across a barrier sweep spanning three orders of magnitude in price
(\Cref{fig:expou-rko}). The forward-volatility agreement fixes today the fair Black
vol of an option starting at a future reset $t_1$; conditional on the field
$\log(S_{t_2}/S_{t_1})$ is Gaussian, so the fair forward vol is a conditional-Black
field average, matching Monte Carlo to within $6.5$\,bp at every reset date
(\Cref{fig:expou-fva}).

On a single Markovian barrier a well-tuned two-dimensional PDE is the fastest route
in our prototypes ($6.6$\,s at benchmark accuracy); an accelerated conditional-Gaussian
pricer (float32 image core, Richardson extrapolation in $M$, a linear-boundary control variate, and
a multilevel-in-grid coarse-image control) reaches Monte-Carlo parity in $9.1$\,s
(\Cref{tab:barrier-timing}). Both are unoptimised Python, so the absolute times and
their ratio are not meaningful beyond order of magnitude, but we expect the ordering
to survive optimisation: a two-factor Markovian barrier is the regime in which
finite differences are most efficient, while the conditional-Gaussian pricer still
integrates over the volatility field, with an inner image-propagator solve per field
node and a residual quadrature error of its own. The field-quadrature route gains instead
through \emph{amortisation} across a book and through \emph{dimension} in the rough
branch, where no low-dimensional PDE competitor exists.

\begin{table}[!b]
\centering
\small
\begin{tabular}{@{}lrrr@{}}
\toprule
method & price & error vs MC (bp) & wall (s) \\
\midrule
Monte Carlo ($2^{18}$ paths)         & $0.51445\ (\pm 9.7\,\text{bp})$ & ---            & $31$ \\
2-D PDE (Craig--Sneyd ADI)           & $0.51491$                        & $+4.6$         & $\mathbf{6.6}$ \\
DS naive (float64 FFT images)        & $0.51611\ (\pm13.2\,\text{bp})$  & $+16.6$        & $127$ \\
DS fast (float32 $+$ Rich.\ $+$ CV)  & $0.51455\ (\pm 8.1\,\text{bp})$  & $+1.0$         & $31$ \\
DS multilevel (coarse-image control) & $0.51452\ (\pm10.7\,\text{bp})$  & $\mathbf{+0.7}$ & $9.1$ \\
\bottomrule
\end{tabular}
\caption{Matched-accuracy wall-clock for the exp-OU upside one-touch at
$B/S_0=1.10$, $\rho=-0.5$ ($\theta=1$, $\sigma=1.8$, $T=1$; single thread). The
acceleration removes the naive propagator's $O(1/M)$ bias \emph{and} its variance;
the multilevel-in-grid variant reaches Monte-Carlo parity in $9.1$\,s, closing to the
2-D PDE, which remains the fastest single-contract route in this Markovian case. All
wall-clock from an unoptimised single-thread Python/NumPy prototype; numerical
precision as labelled per row.}
\label{tab:barrier-timing}
\end{table}

\FloatBarrier

\subsection{Rough exotics on the native field}
\label{sec:rough-exotics}
\begin{figure}[t]
\centering
\includegraphics[width=\textwidth]{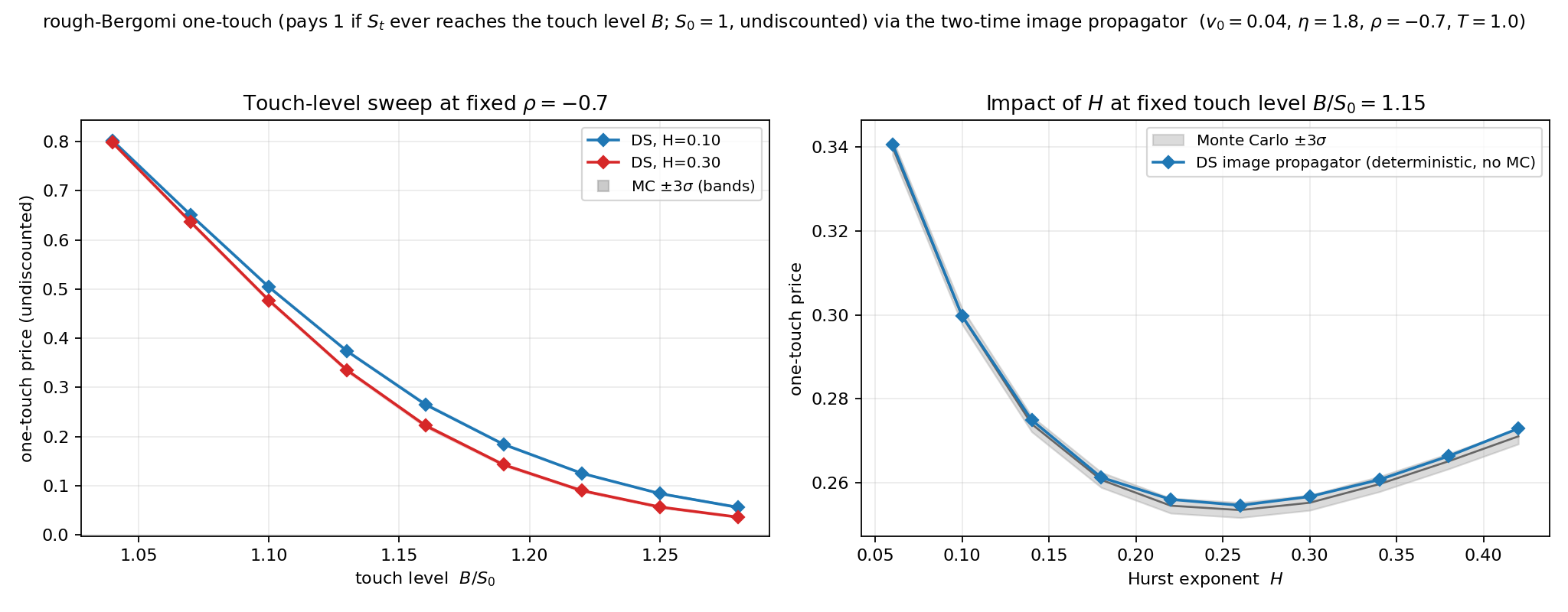}
\caption{Rough-Bergomi upside one-touch via the two-time image propagator
($v_0=0.04$, base vol $\approx20\%$, $\eta=1.8$, $\rho=-0.7$, $T=1$). \emph{Left:}
touch-level sweep at $H\in\{0.1,0.3\}$. \emph{Right:} impact of $H$ at touch level
$B/S_0=1.15$,
non-monotone with an interior minimum. DS (fixed-scramble Sobol field quadrature)
tracks the Monte-Carlo $\pm3\sigma$ bands throughout.}
\label{fig:rb-onetouch}
\end{figure}

\begin{figure}[t]
\centering
\includegraphics[width=\textwidth]{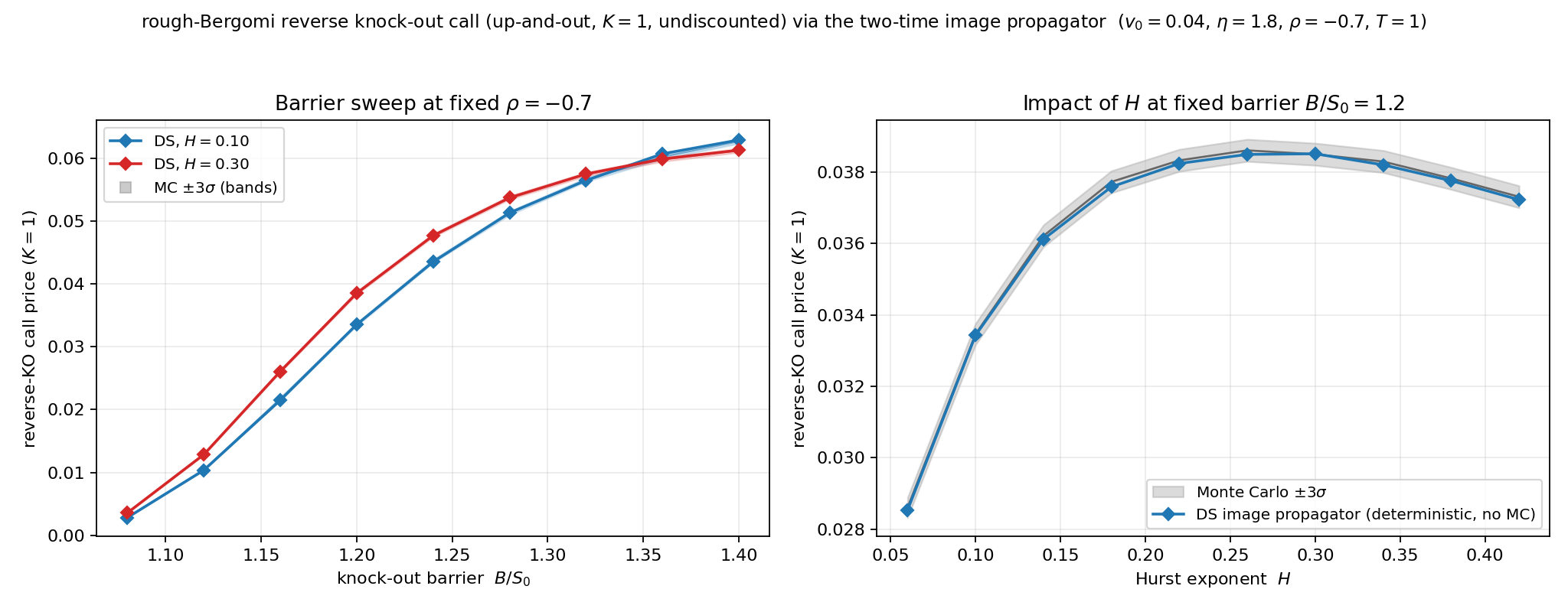}
\caption{Rough-Bergomi reverse knock-out call (up-and-out, $K=1$; $v_0=0.04$,
$\eta=1.8$, $\rho=-0.7$, $T=1$). \emph{Left:} barrier sweep at $H\in\{0.1,0.3\}$, the price
rising with the barrier towards the vanilla-call value.
\emph{Right:} impact of $H$ at $B/S_0=1.2$. DS is
the image propagator with Richardson extrapolation in $M$; bands are Monte-Carlo
$\pm3\sigma$.}
\label{fig:rb-rko}
\end{figure}

\begin{figure}[t]
\centering
\includegraphics[width=0.82\textwidth]{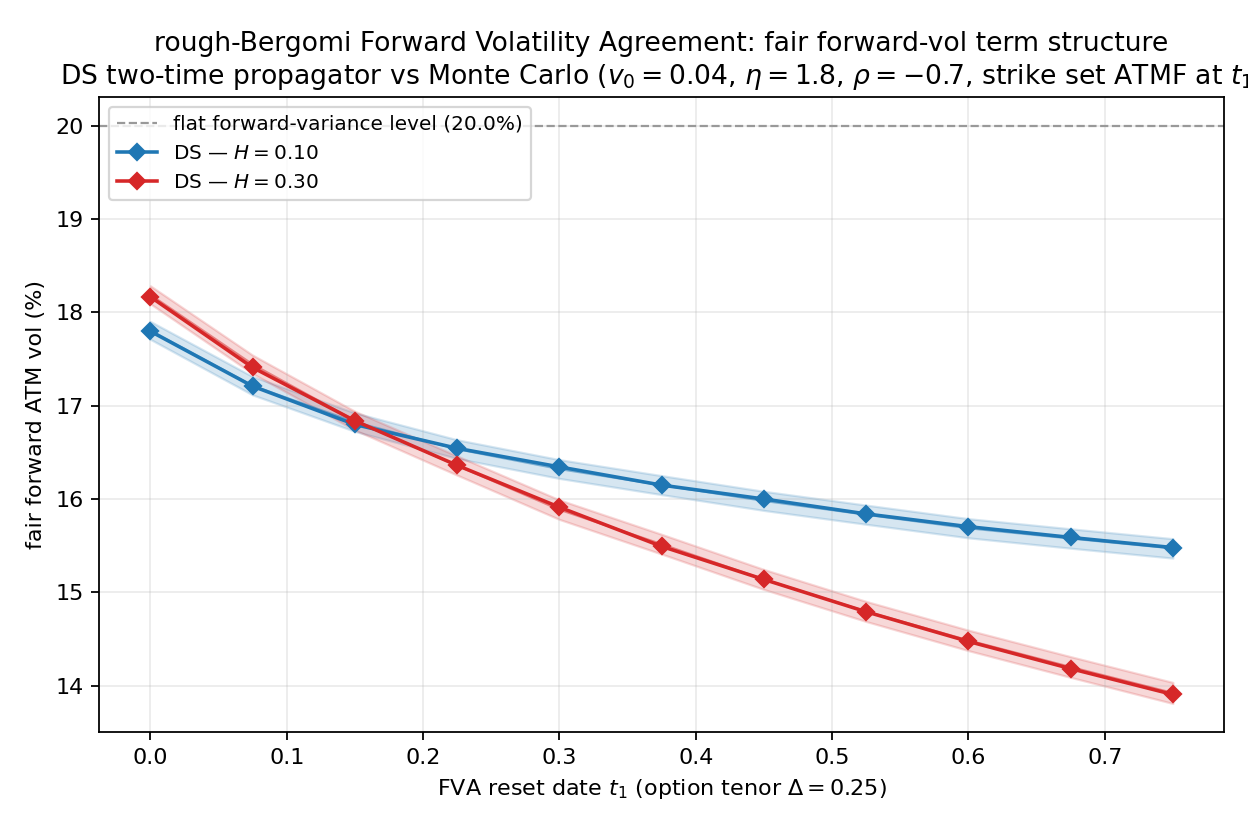}
\caption{Rough-Bergomi forward volatility agreement: fair forward at-the-money vol
versus reset date $t_1$ ($\Delta=0.25$; $v_0=0.04$, $\eta=1.8$, $\rho=-0.7$). The
fair forward vol declines with $t_1$ and $H$ sets the rate, the smoother $H=0.3$
curve falling faster and crossing the rougher $H=0.1$ one. DS matches the Monte-Carlo
$\pm3\sigma$ band to a few basis points throughout.}
\label{fig:rb-fva}
\end{figure}

Passing to rough volatility changes only how the volatility-field paths are
generated. In exp-OU they are drawn from the Markovian OU dynamics; in rough Bergomi
the log-variance on the time grid is a Gaussian vector with the known Volterra
covariance, so each path is produced by mapping a Sobol point through the eigenvectors
of that covariance, ordered by decreasing eigenvalue (PCA-ordered Sobol). Once a field
path is given, the conditional-Gaussian representation \eqref{eq:condgauss}, the image
propagator and the conditional Black formulas are used unchanged. Because
the log-variance field is exactly Gaussian, the full-rank field sample is exact for the
chosen clock. For Europeans and forward starts the pathwise exotic engine then returns
the exact conditional Black value for that discretisation with deterministic field
functionals, so the field average equals the discretised-model price up to the
$O(\Delta t^{H+1/2})$ time discretisation and the outer quadrature, with no
Gaussian-closure contribution; being a positive mixture of arbitrage-free Black prices it
is arbitrage-free at every strike. Barrier survival additionally carries the numerical
image-propagator clock-step error described below. Both this conditional-Black engine and the calibration CF benchmarks of
\Cref{tab:rbergomi} sample the same full-rank discrete field by PCA-ordered Sobol and
impose no rank truncation, but they treat the within-step leverage integral
differently. The exotic engine draws the Brownian driver increments and evaluates the
leverage functional $J$ \emph{pathwise}. For Europeans and forward starts the orthogonal
Brownian component is integrated exactly conditional on the draw, giving the closed-form
conditional-Black value. For barriers its Gaussian transition is exact between clock
points, while survival against the leverage-induced moving boundary is evaluated by the
numerical image propagator, with the clock-step error controlled as described in
\supp{S9}. The CF engine instead samples the discretised field $Y$ and
completes the unresolved within-step increments by the field-first conditional-Gaussian
residual $(m,s^2)$ of \Cref{sec:rbergomi} before COS inversion. The two agree in the
continuum limit but differ at finite clock in exactly this within-step treatment, not
merely in the downstream pricing map. The one-touch (\Cref{fig:rb-onetouch}) shows a non-monotone dependence on
$H$---small $H$ injects short-time bursts that raise the touch probability, trading
off against the terminal variance---and the pricer reproduces the interior turning
point; the reverse knock-out (\Cref{fig:rb-rko}), a boundary-layer functional,
carries a smooth $O(M^{-p})$ bias ($p\approx1.4$) removed by two-point Richardson
extrapolation in the clock count. The forward-volatility agreement transfers directly, and rough
Bergomi's flat forward-variance curve makes the fair forward vol \emph{decline} with
the reset date rather than revert, with $H$ setting the decay rate and the $H=0.3$
and $H=0.1$ term structures crossing---a clean rough-versus-smooth signature in a
plain conditional-Gaussian average (\Cref{fig:rb-fva}).

\begin{figure}[t]
\centering
\includegraphics[width=\textwidth]{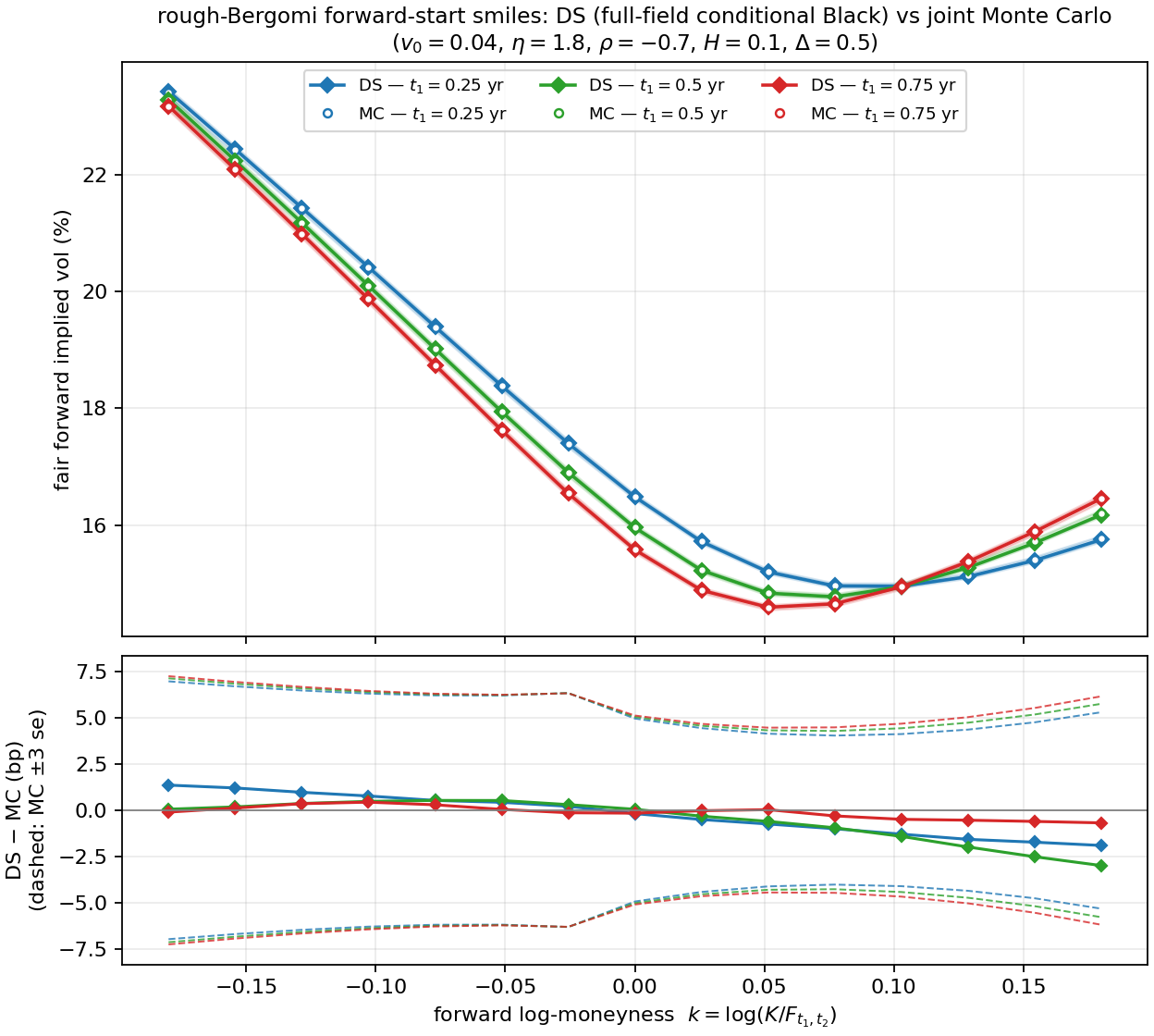}
\caption{Native-rough forward-start smiles in rough Bergomi ($v_0=0.04$, $\eta=1.8$,
$\rho=-0.7$, $H=0.1$, tenor $\Delta=0.5$\,yr) at reset dates
$t_1\in\{0.25,0.5,0.75\}$\,yr against forward log-moneyness $k=\log(K/F_{t_1,t_2})$.
DS is the full-field conditional Black average (no spot simulation); one shared
field prices the whole surface as a field-only quadrature amortised across strikes
and reset dates. Upper panel: implied vols; the Monte-Carlo $\pm3$ standard-error band
is shaded but is thinner than the plotted lines at this scale. Lower panel: DS$-$MC
residual in bp (maximum $3.0$\,bp), with dashed lines marking the $\pm3$ standard-error
band for each reset date.}
\label{fig:fwd-smile}
\end{figure}

The ATMF vol is one strike of a full object. The conditional forward-start price
$C_0(t_1,t_2;k)=\EE_{\text{field}}[\mathrm{Black}_{\mathrm{OTM}}(k;m_{12},q_{12})]$ is
a field average at any $k$, so the entire forward smile at every reset comes from the
\emph{same} field ensemble: the field is built once, after which each additional
strike is a vector operation ($\sim$36\,ms per point when the surface is batched, in
the single-core Python reference).
Across a $45$-point time-zero forward-start surface the maximum DS$-$MC residual is
$3.0$\,bp, every point inside the Monte-Carlo band and within $1.6$ standard errors
(\Cref{fig:fwd-smile}). The single-contract speed-up scales with payoff smoothness
(\Cref{tab:rough-timing}): the one-touch, whose conditional touch probability varies
strongly across paths, gains a modest $1.7\times$, while the smooth forward-vol
payoff, priced in closed form per field draw, is an order of magnitude faster before
any amortisation. Full derivations and the acceleration devices are in \supp{S9}.

The forward smile can also be priced without sampling the field. Conditional on the
field, $\log(S_{t_2}/S_{t_1})$ is Gaussian with mean $-\tfrac12V_{12}+\rho J_{12}$ and
variance $(1-\rho^2)V_{12}$, where $V_{12}$ and $J_{12}$ are the integrated variance and
leverage integral over $[t_1,t_2]$, and the field on that window is Gaussian with the
corresponding block of the two-time covariance. The forward smile is therefore the
vanilla problem on the window, and the composite closure of \Cref{sec:rbergomi} applies
to the block unchanged. On the surface of \Cref{fig:fwd-smile}, against a $2^{21}$-path
simulation with a standard error near $1.9$\,bp, it is within $1.41$, $0.44$ and
$0.64$\,bp RMSE at $t_1=0.25$, $0.5$ and $0.75$ (worst strike $2.0$\,bp), where the
$2^{17}$-point field average is within $1.35$, $0.47$ and $1.84$\,bp, and it takes $0.24$\,s
per reset date on one thread against about $1$\,s for the field average.

\paragraph{Prior art and priority.}
The ingredients are individually standard. Conditioning on the volatility path and
integrating out the orthogonal spot leg analytically is the Romano--Touzi
decomposition for vanillas~\cite{RomanoTouzi1997}, extended to path-dependent payoffs
by Willard~\cite{Willard1997}; volatility-path-conditioned barrier propagation was
developed for classical stochastic volatility by Lipton and Sepp~\cite{LiptonSepp2022},
who solve the inner first-passage problem by heat potentials and observe that a rough
extension should be possible but leave it undeveloped. Rough-volatility exotics can
already be priced by hybrid or affine forward-variance
simulation~\cite{McCrickerdPakkanen2018,Gatheral2022AFV} or by finite-factor Markovian
lifting, and forward-start volatility swaps under rough volatility have short-time
asymptotics~\cite{AlosRolloosShiraya2025}. Our contribution is their combination: a
single engine that prices finite-maturity forward-start smiles and
continuously-monitored touches and knock-outs directly on the native two-time
covariance of a Gaussian--Volterra field, with no spot-path simulation and no Markovian
lift, reducing every product to a field-only quasi-Monte-Carlo average shared across
strikes, barriers and reset dates. The conditioning results supply the inner law;
carrying it over the genuine rough field and demonstrating it under rough Bergomi,
rather than asserting the extension, is the new step.

\begin{table}[t]
\centering
\small
\begin{tabular}{@{}lrrr@{}}
\toprule
rough-Bergomi contract & DS wall (s) & simulation wall (s) & speed-up \\
\midrule
upside one-touch ($B/S_0=1.15$)  & $11$            & $18$  & $\sim\!1.7\times$ \\
forward-vol agreement ($t_1=0.5$)   & $\mathbf{1.6}$  & $22$  & $\mathbf{\sim\!14\times}$ \\
\bottomrule
\end{tabular}
\caption{Single-contract matched-accuracy wall-clock in the rough branch
($v_0=0.04$, $\eta=1.8$, $\rho=-0.7$, $H=0.10$, $T=1$; single thread), where no
low-dimensional PDE competitor exists. The smooth forward-vol payoff, whose inner
option DS prices in closed form, is an order of magnitude faster on a single contract.
The simulation column is a joint Sobol simulation of the rough field and the spot on
the same grid (with a Brownian-bridge barrier correction for the one-touch), run until
its standard error meets the common tolerance ($10^{-3}$ on the touch probability,
$5$\,bp on the fair forward volatility). It is not a conditional (mixing) simulation:
for the forward-volatility agreement the DS route is itself a quasi-random average of
conditional Black prices over the field, so most of the $14\times$ is the spot leg being
integrated analytically, and a conditional simulation would narrow the gap. Wall-clock
from an unoptimised single-thread Python/NumPy prototype (double precision).}
\label{tab:rough-timing}
\end{table}

\FloatBarrier

\section{Fast Risk via the Response Propagator}
\label{sec:risk}
\begin{figure}[t]
\centering
\includegraphics[width=0.86\textwidth]{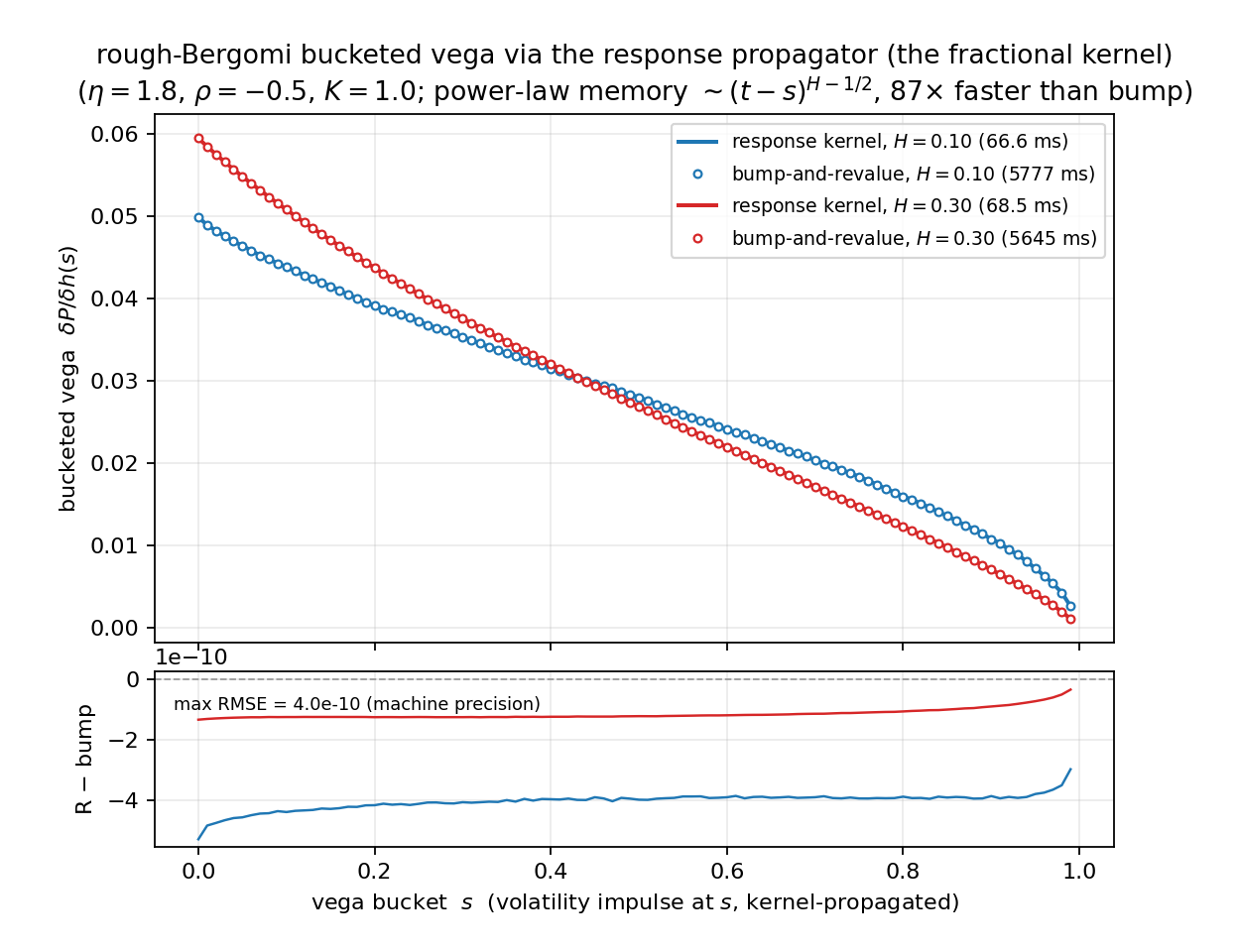}
\caption{Rough-Bergomi bucketed vega via the response propagator (the fractional
kernel; $\eta=1.8$, $\rho=-0.5$, $H\in\{0.1,0.3\}$). The response matrix--vector
product (lines) agrees with common-random-number bump-and-revalue (circles) to
machine precision at $\sim$$80\times$ lower cost (the panel annotates the measured
single-run ratio; timing indicative only); the power-law kernel gives a
long-memory vega profile whose shape is set by $H$.}
\label{fig:greeks}
\end{figure}

The second off-diagonal object, the causal response propagator, carries risk. Spot
$\delta,\Gamma$ are pathwise (matching bump to six digits), and a bucketed-vega term
structure---a latent-factor impulse sensitivity, not a market vega until composed
with the calibration Jacobian---is a single contraction of the payoff sensitivity
against the response: the OU exponential $e^{-\theta(t-s)}$ in the Markovian case,
the fractional kernel $R(t,s)=\eta\sqrt{2H}(t-s)^{H-1/2}$ in the rough case. The
one-pass response reproduces the whole impulse-vega curve in both branches, agreeing
with common-random-number bump-and-revalue to machine precision (RMSE
$\sim10^{-12}$; an automatic-differentiation consistency check of two derivative
implementations, not a pricing-accuracy test) at $\sim$$70\times$ lower cost for the
OU exponential and $\sim$$80\times$ for the fractional kernel, a ratio that in both
cases grows linearly with the number of buckets. The cost ratio is the usual one
between an adjoint (reverse-mode) sweep and one bump per bucket, and we claim no more
than that: any adjoint implementation of the same pricer would achieve it. For a
free Gaussian field the response of the field to a bucket shift is the Volterra
kernel itself, so the shape of the curve is fixed by the model; where the Markovian
response decays exponentially, the fractional kernel decays as a slow power law, and
the vega carries long memory whose shape is set by $H$ (\Cref{fig:greeks}). Chaining the
response with the calibration Jacobian gives market Greeks; the second-order
(differentiated-propagator) equation is written down in \supp{S10} but not tested
here.

\section{Where the Hierarchy Plateaus}
\label{sec:freezing}
Across all models the residual concentrates in the same place: long maturity at the
extremes of correlation for SABR (low $\beta$ with $\rho=-0.9$, and $\rho=0.5$, at
$T=10$); large $\eta$, low $H$, long $T$ for the rough models. The corner limits
expansions around a Gaussian, not the model's prices: the characteristic function itself is
well defined there, and the conditional-mode closures, which condition on the field's
leading modes instead of expanding around them, price it to about a basis point (SABR at
$\beta=0.1$, $\rho=-0.9$, $T=10$ to $1.3$\,bp; rough Bergomi at $H=0.07$, $\rho=-0.9$, $T=2$
to $0.12$\,bp). What the model contributes is the reason every fixed-order skeleton closure
stalls. At $\rho=0$ the characteristic function is the Laplace transform
$\EE[e^{-sI_T}]$, $s=\tfrac12(u^2+\ii u)$, of the integrated variance
$I_T=\int_0^T\sigma_t^2\,\dd t$. For the lognormal-type variance of exp-OU and rough
Bergomi, $I_T$ has all positive polynomial moments but no positive exponential moment,
$\EE[e^{cI_T}]=\infty$ for every $c>0$. Its cumulant generating function
$\log\EE[e^{-sI_T}]$ is therefore not analytic at $s=0$, the cumulant series in $s$ has
zero radius of convergence, and any closure built by truncating that series, or by
adding a fixed number of vertices, is at best asymptotic. The evaluated skeleton
ladder shows this directly (\Cref{tab:ladder}): at strong coupling the levels
alternate in sign, and the three-vertex level improves on the melon level at long
maturities but not at short ones, where the pricing transform needs large $|s|$. Heavier
or rougher fields (larger $\eta$, smaller $H$) move a cell further into this regime. We
have not tested whether the log-variance covariance predicts, cell by cell, where a
given closure falls short of one basis point.

\begin{table}[ht]
\centering
\begin{tabular}{@{}rrr@{}}
\toprule
$\rho$ & $\beta^\ast = 1/(1-\rho^2)$ & admissible damping $\alpha<\beta^\ast-1$ \\
\midrule
$0.0$  & $1.00$ & $<0.00$ \\
$-0.5$ & $1.33$ & $<0.33$ \\
$-0.7$ & $1.96$ & $<0.96$ \\
$-0.9$ & $5.26$ & $<4.26$ \\
\bottomrule
\end{tabular}
\caption{Critical spot moment $\beta^\ast=1/(1-\rho^2)$ and admissible Fourier damping
for rough Bergomi; strongly $\rho$-dependent, in contrast to the $\rho$-free
integrated-variance tail.}
\label{tab:betastar}
\end{table}
\FloatBarrier

Leverage separates the integrated-variance and spot tails. This integrated-variance
heaviness is $\rho$-independent, but the tail of the spot $S_T$ acquires strong leverage
dependence: the critical spot moment is $\beta^\ast=1/(1-\rho^2)$
(\Cref{tab:betastar}), finite for $\beta<\beta^\ast$ and infinite above; for fractional
rough Bergomi with $\rho<0$ this threshold is
sharp~\cite{AndersenPiterbarg2007,Gulisashvili2012,BourdonJeannin2026,Gassiat2019}.
The same critical exponent is visible in the smile: through Roger Lee's moment
formula~\cite{Lee2004} it fixes the asymptotic slope of implied total variance in the
right wing, $\limsup_{k\to\infty}\sigma^2(k)\,T/k=2-4\bigl(\sqrt{p^2+p}-p\bigr)$ with
$p=\beta^\ast-1$.
This is strongly
$\rho$-dependent---$\beta^\ast=1.00,1.33,1.96,5.26$ at $\rho=0,-0.5,-0.7,-0.9$---so
equity leverage \emph{opens} an admissible Fourier strip where there is none at
$\rho=0$: the admissible damping width $\beta^\ast-1$ is $0$, $0.33$, $0.96$ and $4.26$,
growing thirteenfold from $\rho=-0.5$ to $\rho=-0.9$. Contour-deformation pricers are
therefore more viable in the correlated regime than in the uncorrelated case usually
used to demonstrate them. The bound must be imposed
from the formula, not discovered numerically: a Gaussian field average returns
smooth, finite values well past $\beta^\ast$, so the divergence carries no numerical
signature. The practical message is a scope statement: in our tests no fixed-order
skeleton closure reaches one basis point in the near-freezing corner, and the route across
it is to condition on the field, by quadrature over its leading modes or by
quasi-Monte Carlo, rather than to raise the truncation level. Conditioning works there
because it never expands the exponential of the field: the conditional moments are exact
lognormal expectations, and only the law of two scalars given the modes is approximated.
The moment diagnostics are collected in \supp{S7}.

\section{Limitations and Outlook}
\label{sec:limitations}
Every numerical-pricing accuracy result in this article is obtained from controlled
model-to-model validation against a synthetic benchmark of the same model. No
proprietary market quotes are reproduced, and we report no live-market calibrations
here. The distinction is deliberate: a parsimonious pure stochastic-volatility
specification carries fewer parameters than a liquid smile has observations, so a
live-market residual would confound structural misspecification with the numerical
claim the synthetic benchmarks isolate.
What remains between this method and a production tool is therefore narrower than a market
fit---the leverage/Dupire calibration of \Cref{sec:fxhw} and the calibration Jacobian
that turns latent-factor sensitivities into quote-space Greeks. Correctness here is
established numerically rather than analytically: we prove no convergence bounds for the
closures, the conditional-CEV reduction or the displacement booster. The skeleton
ladder of \Cref{sec:diagrams} improves the accuracy by one to two orders of magnitude
per level at moderate coupling, but at strong coupling it is asymptotic rather than
convergent (\Cref{sec:freezing}), and the three-vertex level has been evaluated only
for exp-OU at $\rho=0$. For rough Bergomi the melon level has been run at $\rho=0$
only. The conditional-mode closures are moment closures given the modes: their node law
matches nine moments rather than being exact, their accuracy is established on the grids
reported here, and the jump closure's positivity is checked numerically rather than
guaranteed. They are slower than the fast stacks they replace, about $1$\,s per smile for
rough Bergomi and $8$--$27$\,s for SABR and rough SABR on one thread, against
milliseconds for Hagan's expansion. The SABR routing constants and the jump-closure
settings were chosen on the grid of \Cref{tab:sabr-routing}; the held-out grid checks
them, but a wider out-of-sample test is still needed. The rough barriers are still
priced by quasi-Monte Carlo over the field. The speed
comparisons are against a bump-and-revalue baseline for vega and a joint (not
conditional) simulation for the rough exotics; on a single Markovian barrier a
well-tuned two-dimensional PDE is competitive, and the conditional-Gaussian route is
faster only once the cost is amortised across a book or in the rough branch, where
no low-dimensional PDE competitor exists. The SABR stack is heavier than a single
closed form, and its robustness in production remains to be established.

\section{Conclusion}
\label{sec:conclusion}
The Dyson--Schwinger / 2PI effective-action formalism gives stochastic-volatility
pricing a closure hierarchy that can be evaluated level by level. For the
exponential vertices of lognormal-type variance the skeleton sum is available in
closed form: the melon level reduces the Hartree error on exp-OU by a factor of
$20$--$130$, to $0.01$--$0.11$\,bp across $\rho\in\{0,\pm0.4,-0.7\}$, after an Itô
reduction that makes the correlated vertex local, and it carries over to rough
Bergomi at $\rho=0$ ($0.1$--$1.2$\,bp outside the most extreme cell). The equal-time
Self-Consistent Gaussian (exp-OU at $0.1$--$0.2$\,bp, SABR at a grid mean of
$7.1$\,bp) is a different Gaussian closure, the one of assumed-density filtering. The
locality of the dressed inverse propagator depends on the level: it separates the
Markovian Hartree closures, which reduce to a handful of ODEs, from the rough models
and from the dressed levels, which retain the full two-time propagator. At nonzero
correlation the field is integrated by conditioning on its two leading Gaussian modes,
with exact conditional moments of the composite pair (integrated variance, leverage
integral) and a moment-matched law for that pair. This reproduces rough Bergomi approximately
at the resolution of a $2^{24}$-path reference, about $0.1$\,bp, down to $H=0.07$,
$\rho=-0.9$, and, combined with a Dyson equation
for the correlated drift around the exact absorbed-CEV propagator, resummed by jumps along
the variance clock, SABR to a grid mean of $0.97$\,bp over maturities to ten years and rough SABR to
$0.92$\,bp, deterministically and with non-negative densities in every tested cell; rough
Heston is reproduced to $1$--$5$\,bp. The same two-time objects then supply exotics and
risk: the correlation prices native-rough forward-start smiles, deterministically by the
same closure on the forward window, and continuously monitored barriers as field-only
quadratures, and the causal response gives bucketed vega as one adjoint contraction. Multi-asset consistency is delivered by the FX triangle to
below $2$\,bp.
Calibration to traded quotes with quote-space Greeks, and the freezing-aware
treatment of the near-plateau corner, are left for future work.

\section*{Acknowledgments}
The author thanks Don Wilson, founder and CEO of DRW, for fostering an
environment that encourages research and innovation; Vladimir Sankovich, Head of GQMA,
for valuable discussions and support; and Paolo Lo Presti and Jae Ho Cho for
proofreading the manuscript. Any remaining errors are the author's own.

\medskip
\noindent\textbf{Disclaimer.} The views expressed are the author's own and do not
represent DRW or its affiliates. Nothing herein constitutes investment, legal, or
tax advice, or an offer or solicitation to trade any financial instrument.

\appendix
\section{Map of the Technical Supplement}
\label{app:supp-map}
The accompanying technical supplement (ancillary file on the arXiv abstract page)
provides the derivations and extended evidence supporting the results above, in eleven
sections: \textbf{S1} the doubled-MSRJD 2PI action and its Hartree truncation;
\textbf{S2} the 2PI stationarity equations and the gap-equation reduction;
\textbf{S3} Itô causality and the response-loop structure; \textbf{S4} the complete
exp-OU three-ODE closure; \textbf{S5} the SABR pricing stack (Lamperti drift,
dressed exact-CEV propagator, Islah displacement, booster, boundary-proximity
routing, multi-slice fallback); \textbf{S6} the FX local-vol/stochastic-rate moment
ODEs and the native-rough covariance discretisation; \textbf{S7} extended benchmark
grids, convergence studies and the freezing diagnostics; \textbf{S8} secondary
applications (the quadratic Gaussian Volterra variance model and the FX
triangle); \textbf{S9} exotic-pricing derivations (one-touch,
reverse knock-out, forward-volatility, forward-start, and the acceleration devices);
\textbf{S10} response-equation and Greek details; and \textbf{S11} the conditional-mode
closures (the composite closure for rough Bergomi and its forward window, the jump
closure for SABR and rough SABR, their settings, positivity checks, per-cell results and
cost). Equations, figures and tables
in the supplement are labelled (S.$x$), Figure~S$x$ and Table~S$x$.


\end{document}